# Monte Carlo algorithm based on internal bridging moves for the atomistic simulation of thiophene oligomers and polymers

Flora D. Tsourtou,[1] Stavros D. Peroukidis,[1,2] Loukas D. Peristeras,[3] and Vlasis G. Mavrantzas[1,4,*]

[1]*Department of Chemical Engineering, University of Patras and FORTH-ICE/HT, GR-26504, Patras, Greece*

[2]*School of Science / Technology, Natural Sciences, Hellenic Open University, GR-26335, Patras, Greece*

[3]*National Center for Scientific Research "Demokritos", Institute of Nanoscience and Nanotechnology, Molecular Thermodynamics and Modelling of Materials Laboratory, GR-15310 Aghia Paraskevi Attikis, Greece*

[4]*Particle Technology Laboratory, Department of Mechanical and Process Engineering, ETH-Z, CH-8092 Zürich, Switzerland*

**ABSTRACT**

We introduce a powerful Monte Carlo (MC) algorithm for the atomistic simulation of bulk models of oligo- and poly-thiophenes by redesigning MC moves originally developed for considerably simpler polymer structures and architectures, such as linear and branched polyethylene, to account for the ring structure of the thiophene monomer. Elementary MC moves implemented include bias reptation of an end thiophene ring, flip of an internal thiophene ring, rotation of an end thiophene ring, concerted rotation of three thiophene rings, rigid translation of an entire molecule, rotation of an entire molecule and volume fluctuation. In the implementation of all moves we assume that thiophene ring atoms remain rigid and strictly co-planar; on the other hand, inter-ring torsion and bond bending angles remain fully flexible subject to suitable potential energy functions. Test simulations with the new algorithm of an important thiophene oligomer, α-sexithiophene (α-6T), at a high enough temperature (above its isotropic-to-nematic phase transition) using a new united atom model specifically developed for the purpose of this work provide predictions for the volumetric, conformational and structural properties that are remarkably close to those obtained from detailed atomistic Molecular Dynamics (MD) simulations using an all-atom model. The new algorithm is particularly promising for exploring the rich (and largely unexplored) phase behavior and nanoscale ordering of very long (also more complex) thiophene-based polymers which cannot be addressed by conventional MD methods due to the extremely long relaxation times characterizing chain dynamics in these systems.

*Author to whom correspondence should be addressed.

Phone: +30-2610-997398

Email: vlasis@chemeng.upatras.gr

**1. Introduction**

Poly-thiophenes (PTs) and their parent oligomers, α-conjugated oligothiophenes (OTs), belong to the family of π-conjugated organic materials which find important applications as active semiconducting layers in organic thin film devices.[1-4] The unique combination of optical and electronic properties renders these materials particularly attractive for the design of new, low cost organic based thin film transistors (OTFTs),[5-10] light-emitting diodes (OLEDs)[11,12] and photovoltaic cells.[13,14] Key factors controlling their performance in all these applications are their ordering and morphology at molecular level, because these govern the rate of charge transport through the different layers making up a typical optoelectronic device.[15,16] In particular, the mobility and lifetime of charge carriers and excitons are strongly dependent on molecular factors such as the geometry (dominant molecular conformations) of the molecule, their long-range intra- and inter-molecular ordering and the degree of self-organization.[6,7,17,18] Long-range ordered structures, particularly of lamellae type, facilitate carrier mobility and enhance device performance.

Experimentally, molecular conformation and ordering in conjugated polymers are studied using differential scanning calorimetry (DSC), X-ray diffraction (XRD) and polarized optical microscopy (POM) techniques. With these techniques it has been observed[19, 20] that oligo-thiophenes consisting of up to six thiophene rings (i.e., α-nT with n=2-6) present relatively high melting temperatures and may decompose rapidly while in the melt state.[21, 22] Moreover, it has been demonstrated that α-unsubstituted OTs with up to five thiophene rings do not form mesophases,[19, 21] while the addition of more rings in the chain may result in liquid crystalline behavior at higher temperatures. Nevertheless, at relatively high temperatures decomposition of the chemical compound is possible.[22]

In general, the experimental investigation of the phase behavior of oligo- and poly-thiophenes remains still a challenging task. We mention the XRD, POM and DSC studies of Taliani *et al.*,[23] Destri *et al.*[24] and Porzio *et al.*,[25] which indicated a crystal-to-mesophase transition for α-unsubstituted 6T at the temperature range of 578 K - 585 K, but no mesophase-to-isotropic transition could be detected. According to Destri *et al.*,[24] the mesophase-to-isotropic transition for rigid-rod molecules is hardly detectable due to the large differences in the enthalpies associated with the two processes (5-15 kcal mol$^{-1}$ for the crystal-to-mesophase transition versus 0.1-0.5 kcal mol$^{-1}$ for the mesophase-to-isotropic transition). This is a possible reason why the DSC measurements of Facchetti *et al.*[19] during cooling or heating of short α-unsubstituted nTs (nTs, n=2-6) did not show any second peak that would correspond to a mesophase. Another challenge that Destri *et al.*[24] faced was how to obtain the XRD spectrum of a pure 6T-mesophase, most probably due to the competing condensation reaction that leads to formation of poly-thiophene; thus the mesophase could be assigned neither as a nematic nor as a smectic phase. Additional evidence for the instability of α-unsubstituted thiophenes samples was provided by Kuiper *et al.*,[21] whose DSC measurements upon heating revealed a crystal-to-isotropic phase transition at 540 K for α-5T; however, its decomposition during heating rendered its crystallization unfeasible upon cooling the melt. We note here that for the α-unsubstituted 7T and 8T thiophenes no liquid-crystalline phases have been reported thus far; only their melting points could be measured and were found oddly high.[26, 27] Molecular modeling can help tremendously in this case by addressing questions related with self-organization and nanoscale ordering of these very important materials as they are heated up from their crystal structure or cooled down from their isotropic phase at relatively high temperatures.

Initial attempts to study the structure of thiophene-based oligomers using computer simulations were focused on the crystalline phase. Using molecular mechanics and a modified version of the MM2 forcefield (FF),[28, 29] Ferro et al.[30] investigated the crystal structure of the tetramer (4T), hexamer (6T) and poly-thiophene (PT) of 46-47 thiophene rings,[31] suggesting that the MM2 needs to be improved by assigning partial charges to the atoms of the thiophene rings which should be obtained from quantum-mechanical calculations. Raos et al.[32] carried out a systematic investigation of the inter-ring torsion angle of the 2,2'-bithiophene using *ab initio* calculations with different levels of theory and basis sets. The resulting inter-ring torsion potential was used by Marcon and Raos[33] in a subsequent Molecular Dynamics (MD) simulation study to test different FFs for crystalline OTs, mostly based on the MM3 FF.[34, 35] Although their point-charge model described reasonably well the crystal structure, it did not reproduce accurately the experimental heat of sublimation.[36] It proved, however, an adequate compromise between accuracy and simplicity. In the same spirit, Moreno et al.[37] developed a hybrid model based on the all-atom OPLS FF[38, 39] combined with *ab initio* calculations for the derivation of the atomic charges which was validated in test simulations of thiophene-based oligomers and polymers in the crystalline state. The necessity for revised torsion terms required for the proper description of these materials[40] led to the development of enhanced FF models.[41-43]

Almost all of the previous studies addressed the properties of the crystal structure of oligo- and poly-thiophenes. The first attempt to explore the phase behavior of an unsubstituted thiophene oligomer, α-sexithiophene (α-6T), was reported by Pizzirusso et al.[44] In their study, an all-atom model based on AMBER FF[45, 46] was used, supplemented with additional parameters for partial charges and torsion potentials. Starting from the low-temperature crystal phase, a number of simulations at gradually increasing temperatures were performed. Both the lower-temperature (LT) and the higher-temperature (HT) α-6T polymorphs were predicted correctly. Moreover, three phase transitions were observed and their packing structures were studied in detail: (a) the transition from the LT-crystal to a smectic A (SmA) phase at 577.5 K, (b) the transition from the SmA-to-nematic (Nem) phase at 607.5 K, and (c) eventually the transition from the Nem-to-isotropic (Iso) phase at 667.5 K. We should also mention the work of Zhang *et al.*,[47] who used all-atom MD simulations based on PCFF[48-51] and quantum mechanics calculations to study packing effects on the excitation energies of terthiophene, sexithiophene, and duodecithiophene in the amorphous phase.

It appears that α-6T is the longest unsubstituted OT whose phase behavior has been simulated thus far in full atomistic detail. To study the structure of 20-mer thiophene (20T) system, Curcó and Alemán[52] adopted a generation-relaxation scheme where the generated configurations were relaxed using a conventional Monte Carlo (MC) algorithm. They applied a hybrid five-sited united-atom FF with fixed bond lengths, bond bending and endocyclic torsion angles. The potential governing the inter-ring torsion angle was extracted from a polynomial fit to the quantum mechanical energy profile for the 2,2'-bithiophene case. The van der Waals parameters were adopted from the AMBER FF,[45, 46] while the electrostatic charges were derived from *ab initio* quantum chemistry calculations. The developed FF, combined with a generation-relaxation method, was found to overestimate the density of the amorphous phase by only 3–5%. It also indicated that 20T chains in their amorphous phase tend to be elongated with neighboring thiophene rings belonging to different chains preferring parallel or antiparallel displaced-stacked arrangements. Poly-thiophene has also been studied by Vukmirovic and Wang[53] as a prototype system in order to examine the effect of the absence of side chains on the electronic coupling between backbone chains (compared, for example, to that in poly(3-hexyl thiophene)). The modified CFF91[51, 54] was employed in the simulations with potentials governing inter-ring rotations fitted to *ab initio* calculations.

As the molecular length increases, using atomistic MD simulations in conjunction with an all-atom FF to study phase transitions and structural ordering in poly-thiophenes becomes a formidable task due to the long characteristic times governing molecular relaxation and morphology development in high molecular weight polymers.[55] In this case, coarse-grained (CG) models offer an attractive alternative. The use of a simpler system representation reduces considerably the degrees of freedom describing the system, which helps MD techniques based on CG models to access significantly longer time scales.[41, 42] This approach has been employed in studies of poly(3-alkyl-thiophenes) but also of more complicated systems (e.g., poly(3-hexylthiophene)/C60 blends[41] or ternary systems such as the solvent/poly(3-hexylthiophene)/[6,6]-phenyl-C61-butyric acid methyl ester system[56]). It should be noted that the successful development of CG models relies on the availability of reliable simulation data from all-atom FF simulations or the existence of experimental data;[41, 42] also, their parameterization is valid only within the narrow range of temperature (and pressure) conditions where the effective CG potentials were developed.

Coarse-graining is clearly accompanied by a significant reduction in the degrees of freedom which dramatically reduces computational demands. It has thus gradually gained significant ground in the field. Unfortunately, coarse-graining comes together with several drawbacks:[57, 58] a) It is not clear at all what features of the atomistic model to preserve in the mapping from atomistic units to particles; b) The range of conditions over which reasonable estimates of the structural, thermodynamic, and conformational properties can be expected is usually unknown; c) Restoring the atomistic details lost in the averaging procedure by reverse mapping is a formidable task; d) Dynamic methods at the coarse-grained level should be corrected for the entropy loss (extra dissipation) accompanying the neglected degrees of freedom (otherwise they can be totally erroneous). Indeed, as explained by Öttinger,[58] any level of coarse graining necessarily introduces irreversibility and additional dissipation which is often forgotten or ignored, and this severely limits the usefulness and range of applicability of CG models. To handle the irreversible nature of coarse graining, Öttinger[58] proposed resorting to the fundamental principles of nonequilibrium statistical mechanics, as opposed to strictly at-equilibrium ideas. Given all of the above open and important issues, it is not surprising that, even with the use of some very sophisticated coarse-grained models today, the simulation of structure formation at nanoscale continues to remain an open problem.

An alternative, very promising approach to the problem of the computer simulation of self-organization in complex polymers (such as semiconducting polythiophenes) is offered by the MC method through the use of appropriate moves for the efficient sampling of important configurations. MC relies on the generation of a Markov chain of states, i.e., a sequence of states in which the outcome of a trial move depends only on the state that immediately precedes it.[59-61] Such random states are chosen from a certain distribution $\rho_X(\Gamma)$, where $X$ denotes the macroscopic constraints of the statistical ensemble in which the simulation is carried out and $\Gamma$ the phase space. Transitions are then traced through a scheme that involves: (a) selection of an initial configuration, (b) trial of a new, randomly generated system configuration, and (c) evaluation of an "acceptance criterion" to decide whether the trial configuration will be accepted or rejected. Thanks, in particular, to some very efficient moves tailored for chain systems, equilibration in a MC simulation can be accelerated by 4 to 5 orders of magnitude compared to atomistic MD.[62-66]

So far, the application of the MC method has been restricted to relatively simple systems, such as homopolymers made up of simple repeat units resembling a poly-methylene chain.[62, 66, 67] It has also been applied to simulations of CG (typically one- or two-bead per amino-acid residue) models of polypeptides with simple moves (such as pivot and locally applied concerted rotations), in conjunction with accelerating (replica exchange or Wang-Landau) sampling schemes. But, the capabilities of the method are tremendous. An example that demonstrates its power is that of short alkanethiols self-assembled on metallic substrates.[68, 69] As was shown by Alexiadis *et al.*,[68, 69] through the implementation of a novel MC scheme incorporating a number of sophisticated moves such as atom identity exchange, configurational bias, and intramolecular double rebridging, an initial configuration of alkanethiol molecules randomly placed in the simulation cell above a Au(111) substrate is led to a final state characterized by the formation of a highly ordered and dense monolayer with structural features (tilt, twist, and precession angles) almost identical to those measured experimentally. This is impossible to achieve with atomistic MD.

Motivated by this success of the MC method, in the present work we show how several elementary MC moves already proposed in the literature[63, 65, 70] for the atomistic simulation of simple poly-methylene like molecules (such as linear and branched polyethylene) can be adapted to the more complicated case of thiophene-based oligomers and polymers. Driven by the tremendous scientific and technological interest nowadays in semi-conducting polymers,[2, 71] our long-term objective is to use the new MC algorithm to simulate the liquid crystalline behavior of these materials.[44]

The remainder of this paper is organized as follows: In Section 2 we discuss details of the molecular model used in the new MC algorithm. Section 3 introduces geometric and statistico-mechanical aspects of all MC moves developed in this work for the simulation of oligo- and poly-thiophenes. How the algorithm was parallelized in order to improve its performance is discussed in Section 4. Section 5 presents simulation results from the validation of the new algorithm with an important thiophene-based oligomer, α-sexithiophene (α-6T). It also presents a detailed comparison of the simulation results for the volumetric, structural and conformational properties of this oligomer in its high-temperature isotropic phase with experimental data and a previous simulation study using an all-atom FF. Our paper concludes with Section 6 with a brief summary of the major findings of this work and a short discussion of future plans.

**2. Molecular model and simulation details**

Motivated by the work of Curcó and Alemán[52] and Tiberio *et al.*[72] but also by several other studies[63, 66] in the literature where MC moves are developed for the efficient simulation of polymer systems, in the present MC algorithm we make use of the united-atom representation according to which each C-H group in a poly-thiophene molecule is considered as a single interaction site. Our potential is a *stiff* united-atom model according to which thiophene rings are considered rigid with their atoms lying on the same plane. In addition, all bond lengths are fixed to known values and the same for all intra-ring bond angles (Figure 1). In contrast, inter-ring bond bending and torsion angles are allowed to fluctuate subject to appropriate potential energy functions. These have to be chosen carefully because several previous studies have highlighted the importance of the correct treatment of the potential energies associated with inter-ring torsion angles and non-bonded interactions in accurately predicting the conformational and structural properties of thiophene-based oligomers.[33, 44, 53, 73]

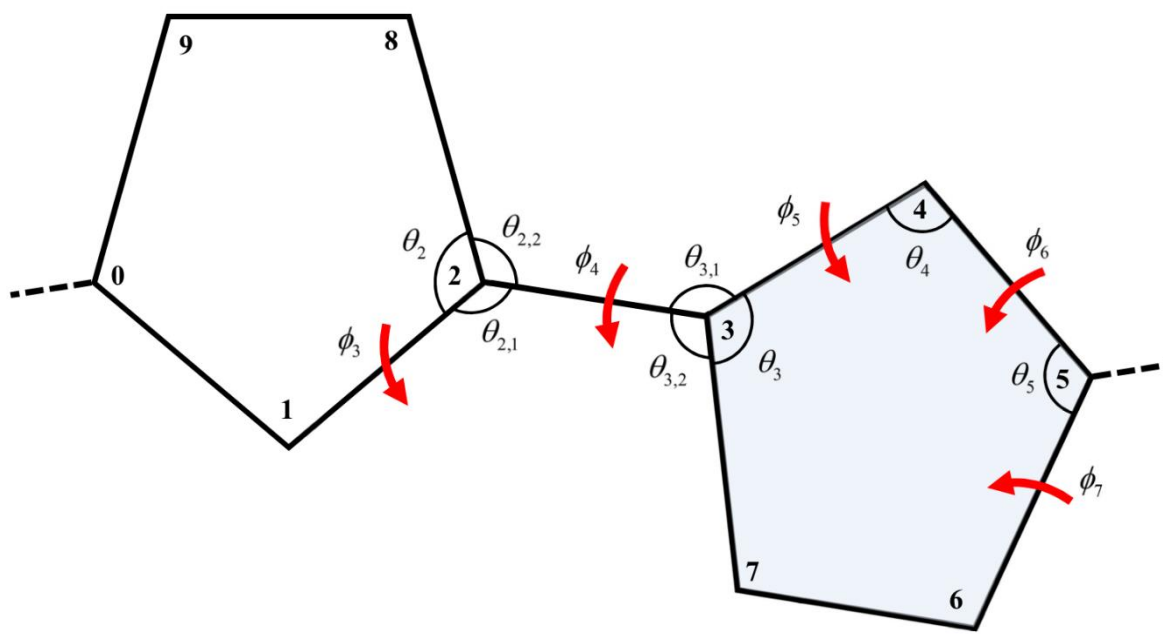

**Figure 1**. Illustration of the detailed geometry of our *stiff* united-atom model around bond 2-3 connecting two successive thiophene rings along a thiophene polymer. Atoms 0, 1, 2, 8, and 9 on the left ring lie on the same plane; similarly, atoms 3, 4, 5, 6, and 7 on the right ring lie on the same plane. Associated with the thiophene ring on the left of the bond 2-3 are the three bond bending angles $\theta_2$, $\theta_{2,1}$ and $\theta_{2,2}$, and the torsion angle $\phi_3$. Similarly, associated with the thiophene ring on the right of the bond 2-3 are the three bond bending angles $\theta_3$, $\theta_{3,1}$ and $\theta_{3,2}$, and the torsion angles $\phi_4$ and $\phi_5$. In our molecular model for MC simulations, all bond lengths are held fixed, and the same holds for all intra-ring bond bending angles on the two thiophene rings (including angles $\theta_2$ and $\theta_3$).

The potential considered here for the inter-ring torsion angle S-C-C-S has been borrowed from the Dreiding FF[69, 74] and is shown by the black curve in Figure 2. It exhibits two local minima, one at $\phi = 0°$ (corresponding to the *cis* conformation) and another at $\phi = 180°$ (corresponding to the *trans* conformation), and a maximum at $\phi = 90°$. The height of the potential at $\phi = 90°$ corresponds to the energy barrier separating the *cis* from the *trans* conformation. The profile of the potential is similar to that proposed by Alberga *et al.*[43] (shown by the green curve in Figure 2) on the basis of Density Functional Theory (DFT) calculations for simulations with poly(3-hexylthiophene) (P3HT) and poly(2,5-bis(3-alkylthiophen-2-yl)thieno[3,2-b]-thiophene) (PBTTT). According to Figure 2, the potential based on Dreiding is characterized by a higher energy barrier at $\phi = 90°$ compared to the Alberga *et al.*[43] potential by 1.07 kcal mol$^{-1}$. This is consistent with recent studies[40, 75] on the torsion potential of PT and P3HT, which suggest that the presence of hexyl chain in P3HT lowers the *cis-trans* barrier relative to PT due to the non-bonded repulsive interaction in the near-planar conformations. Also shown in Figure 2 is the torsion potential used by Pizzirusso *et al.*[44] whose form appears to be slightly different from that of Dreiding in that it exhibits a shallow energy surface around the angle $\phi = 0°$ On the other hand, the two potentials exhibit their local minimum and maximum values at the same values of the angle $\phi$. We also note that the *cis-trans* energy barrier in the torsion potential used by Pizzirusso *et al.*[44] is less by ~ 1.11 kcal mol$^{-1}$ than the energy barrier in the Dreiding torsion potential, implying that the inter-ring torsion potential of Pizzirusso *et al.*[44] is more flexible.

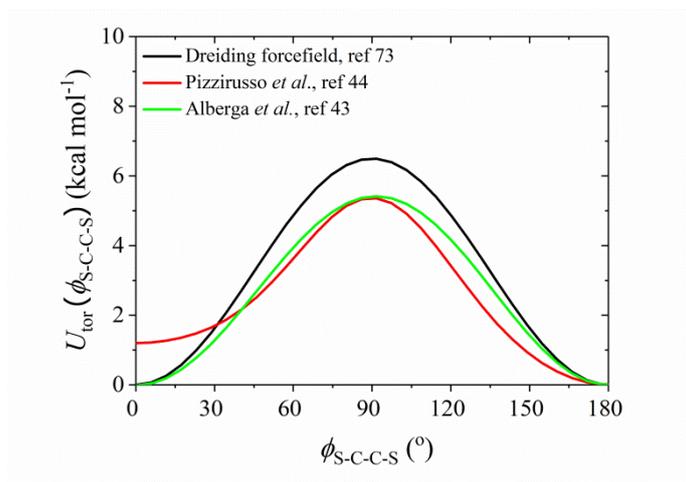

**Figure 2:** Thiophene-thiophene inter-ring (S-C-C-S) torsion angle potential according to the Dreiding forcefield[74] (black curve) used in this work. For comparison we also show the corresponding torsion angle potentials proposed by Alberga *et al.*[43] (green line) and Pizzirusso *et al.*[44] (red line). Only the half of the horizontal axis is shown due to the symmetry of the curves.

In our united-atom model for MC simulations, all non-bonded interactions between sites separated by more than two bonds or belonging to different molecules are described by a classical 12-6 Lennard-Jones (LJ) potential. Moreover, and in order to keep the model as simple as possible (but still maintaining its atomistic nature), electrostatic interactions were neglected and their absence was counter-balanced by appropriately tuning the values of the LJ parameters so that the resulting FF provides volumetric and structural property predictions in the Iso phase in agreement with several data available in the literature[23, 24, 44] at the same thermodynamic conditions. A similar approach (i.e., not to account explicitly for electrostatic interactions) has been followed by Tibiero *et al.*[72] for the prediction of the Nem-to-Iso phase transition temperature of 4-n-alkyl-4'-cyano biphenyls (nCB) series and, in particular, that of 5CB. The computed transition temperature differed only by ~ 5 K compared to the value obtained when Coulomb interactions were explicitly taken into account. Interestingly, in their early study on the prediction of α-4T crystal structure with the MM2 FF[28, 29], Ferro *et al.*[30] had reported that better agreement with experiment is obtained when Coulomb terms are neglected.

An important step in the development of a suitable united-atom model for use in our MC simulations was the parameterization of the LJ parameters describing bonded and non-bonded van der Walls interactions. This was done iteratively as follows. We first chose as a model system a thiophene-based oligomer, α-sexithiophene (α-6T), for which several experimental and simulation data exist in the literature.[23, 24, 44, 76] Then we ran several test MC simulations at high enough temperatures, between 700 K and 740 K (i.e., well above its Nem-to-Iso phase transition)[44] by employing the set of moves described in the next Section to obtain predictions of its volumetric (density), conformational (distribution of torsion angles) and structural (radial pair correlation functions) properties. These were compared with the corresponding sets of data obtained from detailed MD simulations using the all-atom model proposed by Pizzirusso *et al.*[44] and we tuned the epsilon and sigma parameters of the new *stiff* united-atom model for all atomic pairs so that the two sets of predictions (from the all-atom and united-atom models) at a given temperature to coincide.

As a side product of our work, we also developed a *flexible* united-atom model (still without including electrostatic interactions) for use in MD simulations with oligo- and poly-thiophenes in which no geometric constraints are imposed on the molecule. In this *flexible* united-atom model, all bond lengths and bond angles in the thiophene ring can fluctuate. Moreover, thiophene ring atoms are not constrained to lie on the same plane but are left free to fluctuate above and below this plane according to pre-specified bond stretching, bond bending angle and intra-ring torsion angle potentials. For the optimization of the LJ parameters of this *flexible* united-atom model, we followed the same strategy as for the *stiff* united-atom model: namely, we chose them so that MD simulation predictions with the *flexible* model coincided with the all-atom MD simulation predictions of Pizzirusso *et al.*[44] developed on the basis of DFT calculations. We observed that several sets of epsilon and sigma parameters provided good agreement between the two FFs. Thus, a further refinement of the parameters of the *flexible* united-atom model took place by detecting the Iso-to-Nem phase transition point in the MC simulation and tuning the LJ parameters so that this comes as close as possible to the corresponding transition point obtained from the all-atom MD simulations. This will be discussed in a forthcoming publication.

The functional form of all terms describing contributions to the potential energy of the system in the two different united-atom models (the *stiff* version for use in MC simulations and the *flexible* version for use in MD simulations) and their parameters are reported in Tables 1 and 2.

**Table 1.** Functional form of the potential energies of our *stiff* united-atom model for use in the MC simulations and their parameters.

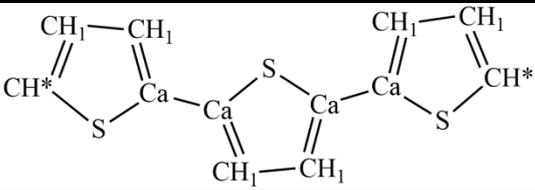

| Atom types | | | |
|---|---|---|---|
| Type of interaction | Molecular model and parameters | | |
| **Non bonded interactions** | $U_{LJ}(r_{ij}) = 4\varepsilon_{ij}\left[\left(\dfrac{\sigma_{ij}}{r_{ij}}\right)^{12} - \left(\dfrac{\sigma_{ij}}{r_{ij}}\right)^{6}\right]$ | | |
| | CH$_1$, CH* | C$_a$ | S |
| $\varepsilon$ (kcal mol$^{-1}$) | 0.08136 | 0.05706 | 0.2064 |
| $\sigma$ (Å) | 3.9994612 | 3.681369 | 3.805741 |
| Mixing Rules | $\sigma_{ij} = \dfrac{\sigma_{ii}+\sigma_{jj}}{2}$ | | $\varepsilon_{ij} = \sqrt{\varepsilon_{ii}\varepsilon_{jj}}$ |

**Bonded Interactions**

| | | | | |
|---|---|---|---|---|
| Bond bending | $U_{\text{bend}}(\theta) = k_{\text{bend}}(\theta - \theta_0)^2$ | *Bending coefficients* | $k_{\text{bend}}$ (kcal rad$^{-2}$) | $\theta_0$ (°) |
| | | S-C$_a$-C$_a$ <br> C$_a$-C$_a$-CH$_1$ | 50 | 120 |

| | | | | | |
|---|---|---|---|---|---|
| Torsion angle | $U_{\text{tor}}(\phi) = \sum\limits_{\text{torsion angles}} k_{\text{tor}}\left[1+\cos(n\phi - d)\right]$ | *Torsion coefficients* | $k_{\text{tor}}$ (kcal mol$^{-1}$) | $n$ | $d$ (°) |
| | | S-C$_a$-C$_a$-S <br> CH$_1$-C$_a$-C$_a$-CH$_1$ <br> CH$_1$-C$_a$-C$_a$-S <br> CH$_1$-CH$_1$-C$_a$-C$_a$ | 3.125 | 2 | -180.0 |
| | | C$_a$-C$_a$-S-C$_a$ | 0.1667 | 2 | -180.0 |

| | CH*-S-$C_a$-$C_a$ |
|---|---|
| | $CH_1$-$CH_1$-$C_a$-$C_a$ |

**Table 2.** Functional form of the potential energies of our *flexible* united-atom model for use in the MD simulations and their parameters.

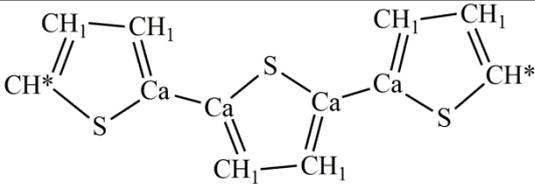

| Atom types | |
|---|---|
| Type of interaction | Molecular model and parameters |

| **Non bonded interactions** | $U_{LJ}(r_{ij}) = 4\varepsilon_{ij}\left[\left(\dfrac{\sigma_{ij}}{r_{ij}}\right)^{12} - \left(\dfrac{\sigma_{ij}}{r_{ij}}\right)^{6}\right]$ | | |
|---|---|---|---|
| | $CH_1$, CH* | $C_a$ | S |
| $\varepsilon$ (kcal mol$^-$) | 0.088140 | 0.061815 | 0.223600 |
| $\sigma$ (Å) | 3.956927 | 3.64664 | 3.769838 |
| Mixing Rules | $\sigma_{ij} = \dfrac{\sigma_{ii}+\sigma_{jj}}{2}$ | | $\varepsilon_{ij} = \sqrt{\varepsilon_{ii}\varepsilon_{jj}}$ |

**Bonded Interactions**

| Bond stretching | $U_{str}(\theta) = k_{str}(l-l_0)^2$ | *Bond coefficients* | $k_{str}$ (kcal Å$^{-2}$) | $l_0$ (Å) |
|---|---|---|---|---|
| | | S-$C_a$<br>S-CH* | 350 | 1.73 |
| | | $C_a$-$C_a$<br>$CH_1$-$CH_1$, -CH*<br>$C_a$-$CH_1$ | 525 | 1.39 |
| Bond | | *Bending coefficients* | $k_{bend}$ (kcal rad$^{-2}$) | $\theta_0$ (°) |

| | | | | | |
|---|---|---|---|---|---|
| bending | $U_{\mathrm{bend}}(\theta) = k_{\mathrm{bend}}(\theta - \theta_0)^2$ | $CH_1$-$C_a$-$C_a$ <br> S-$C_a$-$C_a$ | | | |
| | | $CH^*$-$CH_1$-$CH_1$ <br> $CH_1$-$CH_1$-$C_a$ <br> $CH_1$-$C_a$-S <br> S-$CH^*$-$CH_1$ | 50 | | 120 |
| | | $C_a$-S-$C_a$ <br> $CH^*$-S-$C_a$ | 50 | | 92.4 |
| Torsion angle | $U_{\mathrm{tor}}(\phi) = \sum\limits_{\mathrm{torsion\ angles}} k_{\mathrm{tor}}\left[1 + \cos(n\phi - d)\right]$ | *Torsion coefficients* | $k_{\mathrm{tor}}$ (kcal mol$^{-1}$) | $n$ | $d$ ($^\circ$) |
| | | S-$C_a$-$C_a$-S <br> $CH_1$-$C_a$-$C_a$-$CH_1$ <br> S-$C_a$-$C_a$-$CH_1$ <br> $CH_1$-$CH_1$-$C_a$-$C_a$ <br><br> $CH_1$-$CH_1$-$C_a$-S <br> S-$CH^*$-$CH_1$-$CH_1$ <br> $CH^*$-$CH_1$-$CH_1$-$C_a$ <br><br> $C_a$-$CH_1$-$CH_1$-$C_a$ | 3.125 | 2 | -180.0 |
| | | $C_a$-$C_a$-S-$C_a$ <br> $CH^*$-S-$C_a$-$C_a$ <br><br> $CH_1$-$C_a$-S-$CH_2$ <br> $C_a$-S-$CH^*$-$CH_1$ <br> $CH^*$-$C_a$-S-$C_a$ <br> $C_a$-S-$C_a$-$CH_1$ | 0.1667 | 2 | -180.0 |

Please note the absence of energetic terms associated with improper angles, because preliminary MD simulations with the *flexible* united-atom model developed here showed that these were not necessary for the successful prediction of the mesophase behavior of α-6T (based on the all-atom model by Pizzirusso *et al.*[44]). Also, in all MD and MC simulations, a uniform cutoff of 12 Å was employed for the LJ interactions together with standard tail corrections for the energy and pressure, whereas 1-4 pairwise LJ interactions were taken into account with a factor of 1.

The MC simulations were carried out using three different-sized cubic simulation cells consisting of 140, 560 and 1120 α-6T molecules, respectively, subject to standard periodic boundary conditions along all three space directions. In all cases, the initial configuration consisted of an array of identical α-6T chains in their all-*trans* conformation, placed in a box with density approximately equal to 0.80 g cm$^{-3}$. This configuration was subject to a fast MC relaxation in the *NVT* ensemble at 1000 K to obtain a configuration which is free of excessive long-range atom-atom overlaps, followed by a MC equilibration simulation of almost $10^8$ trial moves (steps) in the *NPT* statistical ensemble at the same temperature and 1 atm. The resulting configuration (with a density approximately equal to 0.72 g cm$^{-3}$) was used as the initial configuration for the subsequent MC simulations in the range of temperatures of interest in this work (between 800 K and 740 K and $P = 1$ atm). In these conditions, the system studied (α-6T) is in the Iso phase and all of its physical quantities of interest were observed to fully equilibrate within about $10^{8}$-$10^{9}$ MC steps (for more details, see Section 4). Test MC simulations at lower temperatures showed that equilibration was slower, especially below 700 K where the first phase transition was observed to occur from the Iso phase to a phase with long range ordering. In all MC runs, the pressure was kept constant, equal to $P = 1$ atm.

The corresponding MD simulations were performed also with a cubic simulation cell containing 1120 α-6T molecules (subject again to full periodic boundary conditions). The initial configuration was created using the Amorphous Builder module[77] integrated in the Scienomics MAPS software,[78] with the mass density set to 0.80 g cm$^{-3}$). The resulting amorphous structure was subjected to a potential energy minimization to remove undesired atom-atom overlaps, followed by a long MD simulation in the *NPT* statistical ensemble at $T = 740$ K and $P = 1$ atm. The duration of the run at these conditions was long enough to ensure full equilibration of the simulated α-6T system at all length scales. The MD simulations were performed with the LAMMPS[79] code using the Nosé-Hoover thermostat-barostat[80, 81] with relaxation times equal to 10 fs and 350 fs, respectively, to maintain the temperature and pressure constant at their desired values. The equations of motions were integrated using the velocity-Verlet method with an integration time step equal to 1.0 fs.

## 3. Monte Carlo moves

The MC algorithm implemented in our work for the simulation of oligo- and poly-thiophenes is based on moves introduced previously for simpler polymer chemistries and architectures such as linear polyethylene,[64, 66] cis-1,4 polyisoprene,[65, 82] H-shaped polyethylene,[83] short- and long-chain branched polyethylene,[83] perfluorinated alkanes,[84] and tri-arm symmetric star polyethylene.[70] It involves only moves which preserve chain connectivity. More complex chain-connectivity altering moves such as end-bridging which typically introduce polydispersity in the system were not implemented.

In the following description of the implemented MC moves, united atoms are denoted by the letter C or S (which stands for carbon and sulfur, respectively) followed by a number which denotes the local index of the specific atom along the backbone of a poly-thiophene molecule. After an attempted MC move, atomistic units in their new positions are denoted by a prime. An example of the terminology used can be seen in Figure 3 illustrating the end-ring rotation move. The move entails the rotation of the very left thiophene ring by changing the inter-ring angle $\phi$ by $\delta\phi$. Then the positions of all atoms C1, C2, S3, C4 and C5 on the rotated ring change and the new ones are denoted respectively as C1′, C2′, S3′, C4′ and C5′.

In the new MC algorithm, most moves involve building a new position for an entire thiophene ring. This is typically achieved by first building three successive atoms of the ring. The positions of the other two atoms then are specified by utilizing the constraint of ring planarity and the fixed values of bond lengths and bond bending angles on the ring. In the following, we will heavily refer to this procedure as the *ring completion* scheme.

(a) **End-ring Rotation:** This move is schematically illustrated in Figure 3 and involves the random rotation of an entire terminal thiophene ring at the beginning or at the end of an α-6T molecule. The move is realized by first specifying the new positions of atoms C4, S3 and C2. First, atom C4′ is built from the positions of atoms C9, S8 and C7 by using the new (chosen) value of torsion angle $\phi_{C4'-C7-S8-C9}$, the known (fixed) value of bond length $l_{C4'-C7} = (l_{C4-C7})$ and by choosing a new bending angle $\theta_{C4'-C7-S8}$; we thus write $C4'(l_{C4'-C7}, \theta_{C4'-C7-S8}, \phi_{C4'-C7-S8-C9})$. Then atom S3′ is built from the known positions of atoms S8, C7 and C4′ by choosing a new value for the torsion angle $\phi_{S3'-C4'-C7-S8}$, a new bending angle $\theta_{S3'-C4'-C7}$ and the known (fixed) value of bond length $l_{C4'-S3'} = (l_{C4-S3})$; we thus write $S3'(l_{C4'-S3'}, \theta_{S3'-C4'-C7}, \phi_{S3'-C4'-C7-S8})$. Finally, atom C2′ is built from the knowledge of the positions of atoms C7, C4′ and S3′ by choosing a new value for the torsion angle $\phi_{C2'-S3'-C4'-C7}$, a new bending angle $\theta_{C2'-S3'-C4'}$ and the known (fixed) value of bond length $l_{C2'-S3'} = (l_{C2-S3})$; we thus write $C2'(l_{C2'-S3'}, \theta_{C2'-S3'-C4'}, \phi_{C2'-S3'-C4'-C7})$. The new torsion angles are selected by sifting the old ones by an angle $\Delta\phi$ randomly selected in the range $(-\Delta\phi_{max}, \Delta\phi_{max})$ with $\Delta\phi_{max} = 60^\circ$. The positions of the remaining two new atoms C1′ and C5′ on the ring are specified by applying the *ring completion* scheme. Using this approach for rotating an end thiophene ring, we sample both its relative rotation and bending with respect to its neighboring ring in the molecule.

Trial values of the non-rigid angles are selected according to the Boltzmann factor of the corresponding bending potential times their sine, i.e., according to $\sin\theta \exp\left[-\beta k_\theta (\theta - \theta_o)^2 / 2\right]$, where $\beta = 1/k_B T$ and $k_B$ is the Boltzmann constant. The corresponding acceptance criterion reads

$$p_{acc} = \min\left[1, \exp(-\beta \Delta U)\right] \qquad (1)$$

where $\Delta U$ is the potential energy difference between initial and final states without accounting for bond bending contributions due to the biased selection of the bond bending angles.

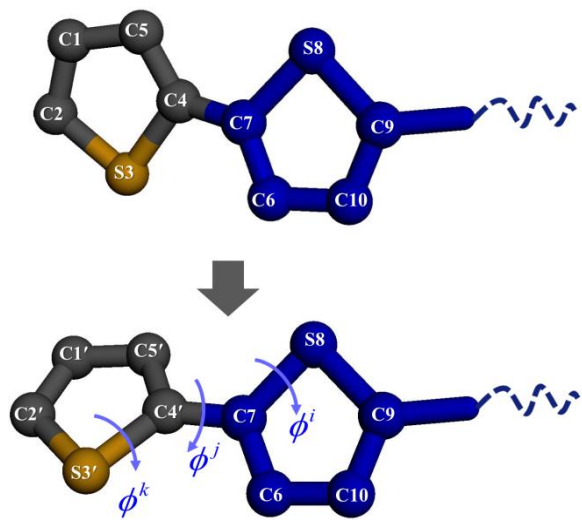

**Figure 3.** Schematic illustration of the end-ring rotation move. Torsion angles $\phi^i$, $\phi^j$ and $\phi^k$ correspond to $\phi_{C4'\text{-}C7\text{-}S8\text{-}C9}$, $\phi_{S3'\text{-}C4'\text{-}C7\text{-}S8}$, and $\phi_{C2'\text{-}S3'\text{-}C4'\text{-}C7}$, respectively.

(b) **Ring-Flip:** In the ring-flip move (depicted in Figure 4), an internal thiophene ring is randomly selected and rotated with respect to its two neighboring thiophene rings on its right and its left by a rotational angle $\omega$ that controls the position of its sulfur atom on its locus, the circle $C^\dagger$ in Figure 4. Initially, the sulfur site S13 is rotated around the axis connecting the two neighboring carbon atoms C12 and C14. Its new position S13′ on the circle $C^\dagger$ defined by the fixed-length bonds $l_{C12\text{-}S13'} = (l_{C12\text{-}S13})$ and $l_{S13'\text{-}C14} = (l_{S13\text{-}C14})$ is specified by the value of the randomly selected rotational angle $\omega$ chosen from a pre-set interval $(-\Delta\omega_{max}, \Delta\omega_{max})$ with $\Delta\omega_{max} = 60^\circ$. Given that two other atoms on the flipped ring remain unchanged (atoms C12 and C14), the positions of the remaining atoms C11′ and C15′ are specified by invoking the *ring completion* scheme. During our implementation of the flip move, the values of four bond bending angles ( C9-C12-S13′, S13′-C14-C17, C9-C12-C11′, C15′-C14-C17 ) and twelve torsion angles ( C10-C9-C12-S13′, C10-C9-C12-C11′, S8-C9-C12-S13′, S8-C9-C12-C11′, C9-C12-C11′-C15′, C9-C12-S13′-C14, C11′-C15′-C14-C17, C12-S13′-C14-C17, S13′-C14-C17-S18, S13′-C14-C17-C16, C15′-C14-C17-S18, C15′-C14-C17-C16 ) change.

In Appendix A we show that the Jacobian entering the acceptance criterion of the ring-flip move depends only on the values of bond lengths $l_{C12\text{-}S13}$ and $l_{S13\text{-}C14}$, and of bond angle $\theta_{C12\text{-}S13\text{-}C14}$; since these remain unaltered in our implementation of the move, the Jacobian cancels out from the acceptance criterion, which finally takes the form:

$$p_{acc} = \min\left[1,\ \exp(-\beta \Delta U)\right] \tag{2}$$

where $\Delta U$ is the potential energy (the sum of bond bending energy, torsion energy and LJ energy) difference between new and old states.

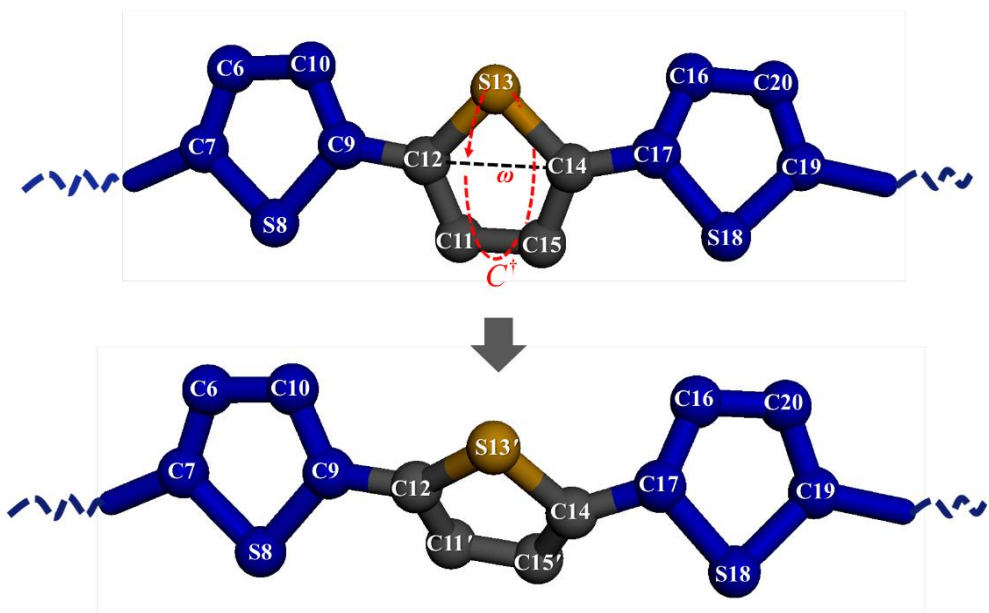

**Figure 4.** Schematic illustration of the ring-flip move.

(c) **Translation:** This move translates the center-of-mass of a randomly selected α-6T molecule and is implemented in two different ways: In the first (which is selected with probability 20%), the molecule is translated along the end-to-end vector of the thiophene molecule by a random distance $dr$ selected from a pre-set interval $(-dr_{max}, dr_{max})$ with $dr_{max} = 1$ Å. In the second (which is selected with probability 80%) the molecule is translated along a vector whose components differ slightly from those of the end-to-end vector by small values randomly selected in the pre-set interval $(-dr_{max}, dr_{max})$ with $dr_{max} = 1$ Å. In both cases, the internal degrees of freedom of the molecule are left unchanged since the molecule is treated as a rigid body, implying also no change in the intramolecular energy of the system. As a result, the acceptance criterion of the move is

$$p_{acc} = \min\left[1,\ \exp(-\beta \Delta U)\right] \tag{3}$$

where $\Delta U$ includes only intermolecular contributions.

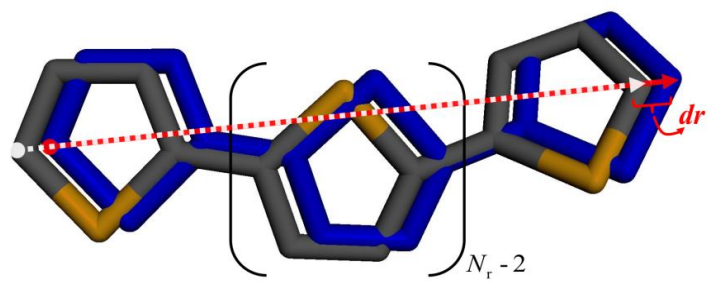

**Figure 5.** Schematic illustration of the translation move. The end-to-end vector of an α-PT molecule is illustrated before (blue chain) and after (grey chain) the trial move with red and white colors, respectively. $N_r$ is equal to the number of thiophene rings in the molecule.

(d) **Rotation:** In this move, a randomly selected poly-thiophene molecule is rotated about an axis of a local orthonormal frame (Oxyz) placed at the center-of-mass of the molecule (see Figure 6). The three axes of Oxyz are defined as follows: axis Oz is taken to be parallel to the end-to-end vector of the selected chain; axis Oy is along the direction of the vector defined by the sum of two vectors that point from the united carbon atoms to the sulfur atom on a randomly selected ring (see Figure 6). The two unit vectors along Oy and Oz are next ortho-normalised using the Gram-Schmidt method which helps define the third axis, axis Ox; this is taken to point in the direction of the vector defined by the cross product of the unit vectors along axes Oy and Oz. The axis (Ox or Oy or Oz) along which the molecule is rotated as a rigid body is chosen randomly. The corresponding angle of rotation $a$ is selected randomly from a pre-set interval $(-a_{max}, a_{max})$ with $a_{max} = 10°$. In the move, the chain is considered as a rigid body, thus its internal degrees of freedom remain unaltered. The same acceptance criterion therefore applies as in the case of the translation move (see eq 3). Our implementation of the chain rotation move is preferable to the standard one where the chain is rotated with respect to an axis of the laboratory frame $Kx'y'z'$ (see Figure 6), because the extended spatial repositioning of the rotated molecule is avoided, thus resulting in higher acceptance rates. As we will discuss in detail in a forthcoming publication, the use of the chain rotation move was essential for observing phase transitions with decreasing temperature, because it helps the system sample efficiently new molecular orientations which are important for obtaining orientationally ordered phases.

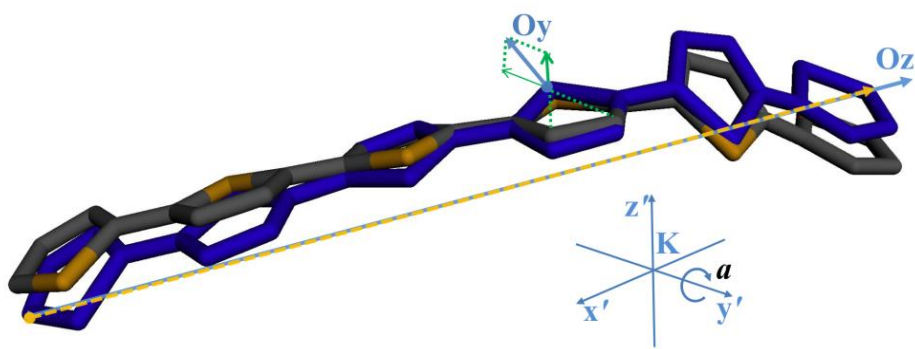

**Figure 6.** Schematic illustration of the rotation move for an α-6T molecule. The dashed line is along the end-to-end vector of the molecule.

(e) **Volume Fluctuation:** This move is used to perform constant pressure simulations by sampling fluctuations in the density of the system. During the move the simulation box is expanded or contracted by modifying all of its edge lengths by a random amount in the interval $(-dr_{max}, dr_{max})$ with $dr_{max} = 0.5$ Å. Then, the first atoms of all chains are affinely displaced following the isotropic deformation of the box while the internal degrees of freedom of each molecule are left unaltered. The acceptance criterion is formulated based on the instantaneous enthalpy difference before and after the move for a system consisting of $N_{ch}$ chains and has the form:

$$p_{acc} = \min\left[1, \exp\left(-\beta(\Delta U + P\Delta V) + N_{ch} \ln \frac{V_{after}}{V_{before}}\right)\right] \quad (4)$$

where $P$ denotes the pressure, $\Delta U$ the intermolecular potential energy difference before and after the move, $V_{before}$ the volume before the move and $V_{after}$ the volume after the move.

(f) **Bias-Reptation:** During this move, an end thiophene ring from a randomly selected molecule is cut off and appended to the other end of the same chain (see Figure 7). Given the need for a position with a relatively large free volume where the reptated ring can be accommodated, the torsional degree of freedom (torsion angle) defining the new position of the ring should be selected in a biased fashion.[85] Thus, following the standard practice[85] the selected end ring is cut off and the excised segment C2-S3-C4 ($N_{cut} = 3$) is rebuilt on the other end of the same poly-thiophene molecule by selecting a number of trial positions $N_{trial}$ for each atom to be regrown; six possible trial positions are generated ($N_{trial} = 6$) for each atom in our implementation. As in the case of the end-rotation move, the two bending angles $\theta_{Sk3\text{-}Ck4\text{-}C2'}$ and $\theta_{Ck4\text{-}C2'\text{-}S3'}$ defining the position of the trimer are selected based on the corresponding Boltzmann weights of their bong angle energies. In addition, each of the three torsion angles $\phi_{Ck2\text{-}Sk3\text{-}Ck4\text{-}C2'}$, $\phi_{Sk3\text{-}Ck4\text{-}C2'\text{-}S3'}$ and $\phi_{Ck4\text{-}C2'\text{-}S3'\text{-}C4'}$ needed to define the new positions of atoms $C2'$, $S3'$ and $C4'$, respectively, are picked up from the $N_{trial}$ angles with probability given by:

$$p_i^{old \to new} = \frac{\exp(-\beta U_{i,tor}^j)}{\sum_{j=1}^{N_{trial}} \exp(-\beta U_{i,tor}^j)} \quad (5)$$

where $U_{i,tor}^j$ denotes the torsion energy corresponding to the case that atom $i$ is placed in trial position $j$. The bias introduced by the selection process is removed by using the Rosenbluth statistical weight of the new state:

$$W^{old \to new} = \prod_{i=1}^{N_{cut}} \left[\sum_{j=1}^{N_{trial}} \exp(-\beta U_{i,tor}^j)\right] \quad (6)$$

To satisfy the criterion of microscopic reversibility, the statistical weight of the old state $W^{old \to new}$ associated with the inverse problem should also be calculated in the same fashion. The ring reconstruction is finally completed by applying the *ring completion* scheme to specify the positions of the remaining two new atoms C1′ and C5′. The acceptance criterion of the bias-reptation move is

$$p_{acc} = \min\left[1, \frac{W^{old \to new}}{W^{new \to old}} \exp(-\beta \Delta U)\right] \quad (7)$$

where $\Delta U$ denotes the potential energy difference between initial and final states without accounting for bond bending and torsion angle contributions.

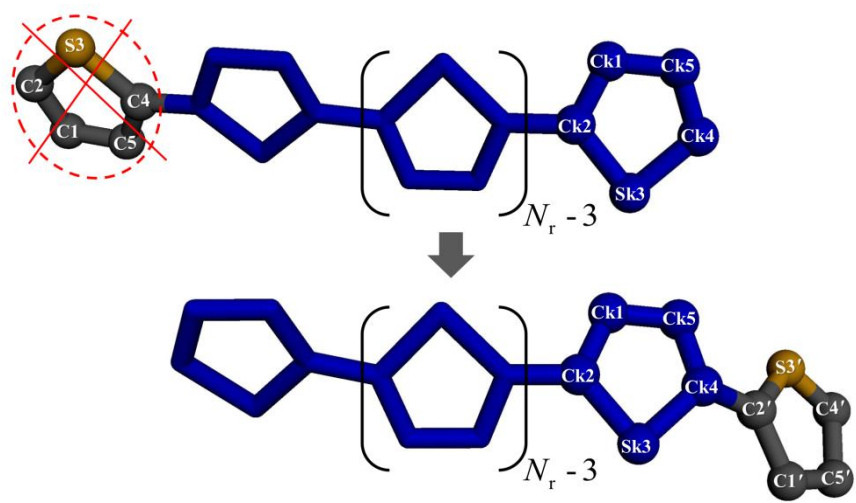

**Figure 7.** Schematic illustration of the bias-reptation move. $N_r$ is equal to the number of thiophene rings in the molecule and k5 ($= 5N_r$) is the total number of atoms in a chain.

(g) **Concerted Rotation**: This move (ConRot) was introduced by Dodd *et al.*[86] in order to rearrange internal parts of a polymer chain in a concerted motion fashion that alters several internal degrees of freedom in a coordinated way, while maintaining chain integrity. In our implementation for poly-thiophene molecules, this move is initiated by rearranging a C-S-C trimer in a randomly chosen thiophene ring, e.g., segment C12-S13-C14 in Figure 8. The two atoms C9 and C17 neighboring the trimer are displaced through a small rotation of the corresponding rings around the bonds $l_{C7-S8}$ and $l_{S18-C19}$, respectively. This alters also the positions of atoms C6, C10, C16 and C20. The corresponding rotation angles ($\omega'$ and $\omega''$ in Figure 8) are randomly selected within pre-specified bounds $(-\Delta\omega_{max}, \Delta\omega_{max})$ with $\Delta\omega_{max} = 10°$. The driving of the neighboring atoms leads to improved sampling of the trimer position, which is determined by solving the geometric problem of trimer bridging.[66, 86, 87] The new positions C12′, S13′ and C14′ are defined by specifying the values of four bond lengths ($l_{C9'-C12'}$, $l_{C12'-S13'}$, $l_{S13'-C14'}$ and $l_{C14'-C17'}$) which in our case are fixed and of five bond angles ($\theta_{S8-C9'-C12'}$, $\theta_{C9'-C12'-S13'}$, $\theta_{C12'-S13'-C14'}$, $\theta_{S13'-C14'-C17'}$ and $\theta_{C14'-C17'-S18}$). In our implementation of the ConRot move, these five bond angles are selected in a bias way according to the Boltzmann factors of their corresponding bond bending potentials, exactly as we did in the case of the end-ring rotation move. Using these predefined values, the solution of the bridging problem provides a maximum of 16 possible solutions for the trimer that reconnects the driven atoms on its left and right. Of these, only a small subset of $N_{sol}^{new}$ results in reasonable torsion angles (i.e., inter-ring S-C-C-S torsions close to *cis* or *trans* conformation) and no overlaps with neighboring atoms. From this subset, a random solution $i_{sol}^{new}$ is selected based on the Boltzmann factor of the corresponding torsion energy $U_{tor(i_{sol}^{new})}$. Therefore, the attempt probability $p_{select}^{old \to new}$ for the forward move is given by:

$$p_{select}^{old \to new} = \frac{\exp\left[-\beta U_{tor(i_{sol}^{new})}\right]}{\sum_{i=1}^{N_{sol}^{new}} \exp\left[-\beta U_{tor(i_{sol}^{new})}\right]} \tag{8}$$

Finally, the positions of new atoms C11′ and C15′ are computed according to the *ring completion* scheme mentioned several times so far. During the ConRot move, the configuration of the chain is drastically modified, because eleven inner atoms C6, C9, C10, C11, C12, S13, C14, C15, C16, C17 and C20 are repositioned. The microscopic reversibility is imposed by attempting the reverse move as well, which results in the original chain configuration; the corresponding attempt probability is $p_{select}^{old \to new}$. The formulation of the trimer bridge solution requires the transformation from Cartesian coordinates to internal coordinates, thus the corresponding Jacobian of the transformation[66] should also be used in the Metropolis criterion of the move, which is eventually expressed as

$$p_{acc} = \min\left[1, \ \frac{p_{select}^{new \to old} J^{new}}{p_{select}^{old \to new} J^{old}} \exp(-\beta \Delta U)\right] \quad (9)$$

where $\Delta U$ denotes the potential energy difference between initial and final states without taking into account contributions due to the chosen biased bond bending angles and $J^{new}$ and $J^{old}$ are the respective Jacobians. Their detailed mathematical form is reported in refs 66, 86 and 87.

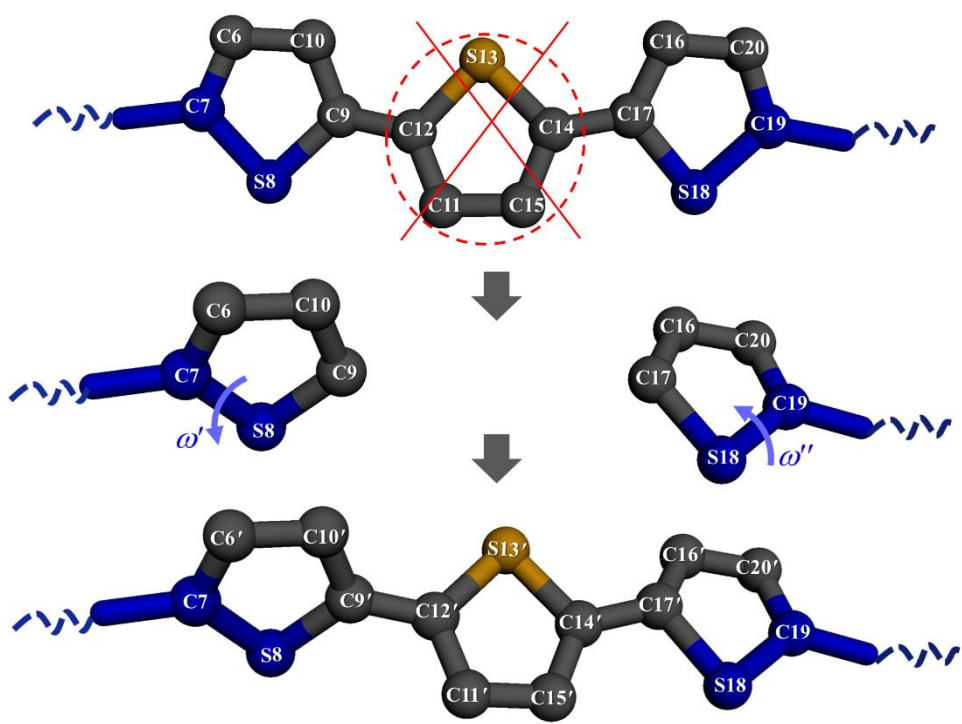

**Figure 8.** Schematic illustration of the ring-ConRot move.

## 4. Moves acceptance rates - Parallelization of the new Monte Carlo algorithm

We carried our preliminary MC runs in the *NPT* ensemble at $P = 1$ atm and at several temperatures above 700 K for the system to be in its isotropic state using the following mix of moves which gave almost optimal efficiency: 20 % bias-reptations, 7 % end-ring rotations, 5 % ring-flips, 10 % ConRot's, 25 % translations, 30 % rotations and 3 % volume fluctuations. As already mentioned in Section 2, we worked with three different systems, containing 114, 560 and 1120 α-6T molecules of almost the same initial density (0.72 g cm$^{-3}$). The resulting acceptance rates of the seven MC moves and the total number of MC iterations needed to fully equilibrate the systems as functions of system size and temperature (in the region between 700 K and 800 K) are reported in Table 3. The total number of iterations for complete equilibration was computed based on the time needed for the time autocorrelation function of the unit vector along the end-to-end vector of the α-6T molecule to drop to zero (see also Section 5.2 below). The very low acceptance rates of the bias-reptation and ConRot moves are evident for all systems: only about 0.30 % of the attempted bias-reptation and only 0.12 % of the attempted ConRot moves are accepted. Despite this, however, our proposed MC algorithm is capable of fully equilibrating the three tested systems within reasonable CPU time. It is also interesting that the acceptance rates of the moves are practically system size - independent. Obviously, as the temperature is lowered, the acceptance rates of all moves drops, which might be a problem when the algorithm is used to simulate morphological transitions at lower temperatures associated with the appearance of ordered phases. The problem can be overcome by implementing a parallel version of the new algorithm, very similar in nature to the parallel tempering method.[82] According to this, an extended ensemble of systems is considered each one of which is a replica of the original system, equilibrated however at different (smaller) values of the repulsive core diameters characterizing van der Waals interactions in the system, at the same temperature. In this parallel MC method, except from the MC moves described in Section 3, an extra move is introduced, realized by swapping configurations between systems with adjacent Lennard-Jones core diameters. The method has already been used with great success by Alexiadis *et al.*[69] in simulating the self-assembly of alkane-thiols on Au(111) at surface densities close to full coverage.

**Table 3**. Attempt probabilities and average acceptance rates of the MC moves employed in the new Monte Carlo algorithm and total number of MC iterations needed to fully equilibrate a bulk isotropic α-6T system. Results are reported as a function of system size and temperature.

| Move | Attempt | Acceptance rate % |
|---|---|---|

|  | probability % | 560-chain system | | 1120-chain system |
|---|---|---|---|---|
|  |  | 740K | 800K | 800K |
| end-ring rotations | 7 | $5.20 \pm 0.30$ | $6.00 \pm 0.2$ | $6.00 \pm 0.3$ |
| bias-reptations | 20 | $0.20 \pm 0.04$ | $0.30 \pm 0.03$ | $0.30 \pm 0.03$ |
| Translations | 25 | $12.10 \pm 0.50$ | $13.50 \pm 0.3$ | $13.70 \pm 0.3$ |
| Rotations | 30 | $27.00 \pm 0.40$ | $29.00 \pm 0.3$ | $29.10 \pm 0.3$ |
| ring-flips | 5 | $17.50 \pm 0.30$ | $18.40 \pm 0.3$ | $18.50 \pm 0.3$ |
| volume fluctuations | 3 | $3.10 \pm 0.20$ | $3.20 \pm 0.3$ | $2.00 \pm 0.3$ |
| ConRots | 10 | $0.12 \pm 0.05$ | $0.14 \pm 0.05$ | $0.12 \pm 0.05$ |
| **Monte Carlo iterations** |  | $1 \times 10^9$ | $5 \times 10^8$ | $7 \times 10^8$ |

The average CPU time per MC iteration for serial execution was calculated to be approximately 0.013 seconds. Therefore, a MC simulation of $10^9$ steps requires a significant execution wall time, on the order of many days. To speed up algorithm execution, we profiled the new code and as expected[84] we found that volume fluctuation was by far the most computational intensive part of the algorithm due to the need to re-calculate (whenever an attempted volume fluctuation move passed the early rejection criteria) and update (whenever the move was accepted) the Verlet neighbor list. Even in cases when volume fluctuation was attempted with as low a probability as 1 %, its cost (fraction of total execution time spent in this move) could be as high as 78 % (see Table 3). To overcome this, the volume fluctuation move was parallelized using two different techniques, one based on Message Passing Interface (MPI) and another based on GPU multithreading. The MPI implementation made use of the force decomposition approach. During the attempt phase of a volume fluctuation move, each process calculates the new positions of the atoms for a number of α-6T chains (the number of chains is equally distributed among the processes). Then, the coordinates of all atoms in the system are updated across the processes using the collective communication functionality of MPI, with the number of pairs to be checked for the update of the various lists (Verlet, linked cells, etc.) being equally distributed among them. As far as the GPU implementation is concerned, the interested reader is referred to a recent work by our group (see ref 84).

The improvement in computational efficiency for each of the two techniques was quantified by calculating the speedup $S$ defined as the ratio of wall time for serial execution over wall time for parallel execution. Although the achieved speedup depends on a number of factors such as the size of the system, the percentage of attempting a volume fluctuation move and the CPU/GPU device combination, the GPU implementation gave always a higher performance. Our benchmark tests were performed on an Intel(R) Xeon(R) E5-2660v3 (2.6GHz) CPU, combined with an NVIDIA Tesla K40 GPU device using a simulation cell containing 560 α-6T chains (corresponding to a total of 16,800 atoms) at temperature $T = 740$ K and pressure $P = 1$ atm with the same mix of moves as the one reported in Table 3, but by progressively increasing the percentage of volume fluctuations at the expense of ring-flips and end-ring rotation moves. The results obtained are presented in Table 4. We see that the overall speedup $S$ achieved was close to 6 for the GPU implementation and close to 3 for the MPI implementation. Therefore, the GPU implementation was preferred in all subsequent MC simulations in this work with the new MC algorithm.

**Table 4**. % attempt probability of volume fluctuation moves, their average computational cost ($C_{VF}$) and the corresponding speedups $S$ achieved with the MPI and GPU implementations. The simulation was performed with 560 α-6T chains in a cubic periodic box with initial dimensions $L_x = L_y = L_z = 80$ Å. Here, we report the optimum MPI results achieved using 12 CPUs. In all cases, the error in the speedup was equal to 0.05.

| % attempt probability of volume fluctuation moves | $C_{VF}$ (%) | $S_{MPI}$ | $S_{GPU}$ |
|---|---|---|---|
| 1 | 78 | 2.86 | 3.41 |
| 3 | 93 | 2.86 | 6.02 |
| 5 | 96 | 3.49 | 6.99 |

## 5. Results and Discussion

**5.1. Validation.** To check that all moves were properly implemented in the new MC code, we first tested the algorithm in simulations with α-6T phantom chains (that is α-6T chains with only bond angle bending interactions) at $T = 1000$ K. To check each move separately, several MC tests were made with different mixes of moves. For example, to check the proper implementation of the bias reptation move, we ran phantom chain MC simulations with 100% bias reptations. To check the flip move, on the other hand, we ran simulations with only 2% bias reptation moves and 98% flips (because with the flip move alone the code is not ergodic). And the same with the end-ring rotation and ConRot moves. From the phantom chain simulations, one calculates the normalized distribution of bond bending and torsion angles, which can then be compared with the corresponding theoretical distributions from the same FF.

For poly-thiophene molecules, of interest in our phantom chain MC simulations are the distributions of the four bond angles $\theta_{2,1}$, $\theta_{2,2}$, $\theta_{3,1}$ and $\theta_{3,2}$, and of the two inter-ring torsion angles $\phi_3$ and $\phi_4$ around the bond connecting two successive thiophene rings. All of them are defined in Figure 1 presenting the detailed geometry locally along the bond connecting two successive thiophne rings. The torsion angle $\phi_3$ refers to atoms 0, 1, 2 and 3 in Figure 1 and is known as the C-S-C-C inter-ring torsion angle. Similarly, the torsion angle $\phi_4$ relates to atoms 1, 2, 3 and 4 in Figure 1 and is known as the S-C-C-S inter-ring torsion angle. The theoretical distributions of these four bending and two torsion angles are derived in Appendix B. We first calculated the probability distribution $P(\theta_{2,1}, \theta_{2,2}, \theta_{3,1}, \theta_{3,2})$ of the bond bending angle quad $(\theta_{2,1}, \theta_{2,2}, \theta_{3,1}, \theta_{3,2})$. We found (see Appendix B.1):

$$P(\theta_{2,1}, \theta_{2,2}, \theta_{3,1}, \theta_{3,2}) \propto$$

$$\frac{\left(\prod_{i=2}^{3}\prod_{j=1}^{2}\exp\left[-\beta U_{\text{bend}}(\theta_{i,j})\right]\right)}{\sqrt{1-\cos^2\theta_2-\cos^2\theta_{2,1}-\cos^2\theta_{2,2}+2\cos\theta_2\cos\theta_{2,1}\cos\theta_{2,2}}}$$
$$\frac{\sin\theta_{3,1}\sin\theta_{3,2}}{\sqrt{1-\cos^2\theta_3-\cos^2\theta_{3,1}-\cos^2\theta_{3,2}+2\cos\theta_3\cos\theta_{3,1}\cos\theta_{3,2}}}d\theta_{2,1}d\theta_{2,2}d\theta_{3,1}d\theta_{3,2}d\varphi_4 \quad (10)$$

This implies that

$$P(\theta_{2,1}, \theta_{2,2}, \theta_{3,1}, \theta_{3,2}) \propto P(\theta_{2,1}, \theta_{2,2})\, P(\theta_{3,1}, \theta_{3,2}) \quad (11)$$

i.e., the probability distribution functions of the pairs $(\theta_{2,1}, \theta_{2,2})$ and $(\theta_{3,1}, \theta_{3,2})$ are independent and mathematically identical, because of the equivalence of stiff angles $\theta_2$ and $\theta_3$ (see Figure 1). Moreover, $P(\theta_{2,1}, \theta_{2,2})$ is symmetric with respect to $\theta_{2,1}$ and $\theta_{2,2}$, and $P(\theta_{3,1}, \theta_{3,2})$ symmetric with respect to $\theta_{3,1}$ and $\theta_{3,2}$. That is, the probability distributions for all four angles $\theta_{2,1}$, $\theta_{2,2}$, $\theta_{3,1}$ and $\theta_{3,2}$ are the same!

By further moving to a description in terms of the angle $\theta_{2,2}$ from a description in terms of the torsion angle $\phi_3$ (the torsion angle of atoms 0, 1, 2 and 3 in Figure 1), calculating the corresponding Jacobian and substituting back (see Appendix B.2), we find:

$$P(\theta_{2,1}, \theta_{2,2}) = d\theta_{2,1}d\theta_{2,2}\int d\theta_{3,1}d\theta_{3,2}\, P(\theta_{2,1}, \theta_{2,2}, \theta_{3,1}, \theta_{3,2})$$
$$\propto \frac{\sin\theta_{2,1}\sin\theta_{2,2}\exp\left[-\beta U_{\text{bend}}(\theta_{2,1})\right]\exp\left[-\beta U_{\text{bend}}(\theta_{2,2})\right]}{\sqrt{1-\cos^2\theta_2-\cos^2\theta_{2,1}-\cos^2\theta_{2,2}+2\cos\theta_2\cos\theta_{2,1}\cos\theta_{2,2}}}d\theta_{2,1}d\theta_{2,2} \quad (12)$$

Therefore, the normalized probability distribution of $\theta_{2,1}$ is given by the following expression:

$$\rho(\theta_{2,1}) = \frac{\int_0^\pi d\theta_{2,2} P(\theta_{2,1}, \theta_{2,2})}{\int_0^\pi d\theta_{2,1} \int_0^\pi d\theta_{2,2} P(\theta_{2,1}, \theta_{2,2})} \qquad (13)$$

and the same expression holds for $\theta_{2,2}$, $\theta_{3,1}$ and $\theta_{3,2}$. The integral appearing in eq 13 can be easily calculated using MC integration and the result (the same for the four angles) is shown in Figure 9, together with the corresponding distribution as computed directly from our MC algorithm with phantom α-6T molecules using 100% bias-reptation moves. The two sets of data coincide, which is an indirect proof that the bias-reptation move was properly implemented in our MC algorithm. As explained above, the test was repeated with different mixes of MC moves, each time giving the highest weight to the move whose implementation we wanted to check. For example, to check the correctness of the implemented ring-flip move, the mix of moves used consisted of 1% bias-reptations, 1% ConRot's and 98% ring-flips, i.e., we included a small percentage of bias-reptations and ConRot's in order to render the simulation ergodic (because the way the ring-flip was implemented in our code does not sample degrees of freedom associated with the two end thiophene rings). Again, simulated and theoretical probability distributions were found to coincide, implying that also the ring-flip move was properly implemented. And the same procedure (to apply 1% bias-reptations) was followed to check that also the ring ConRot move was properly implemented. We further note here that the rigid chain translation, the rigid rotation and the volume fluctuation moves do not affect the internal degrees of freedom of the simulated molecule, thus no phantom chain MC tests could be made to check their correct implementation.

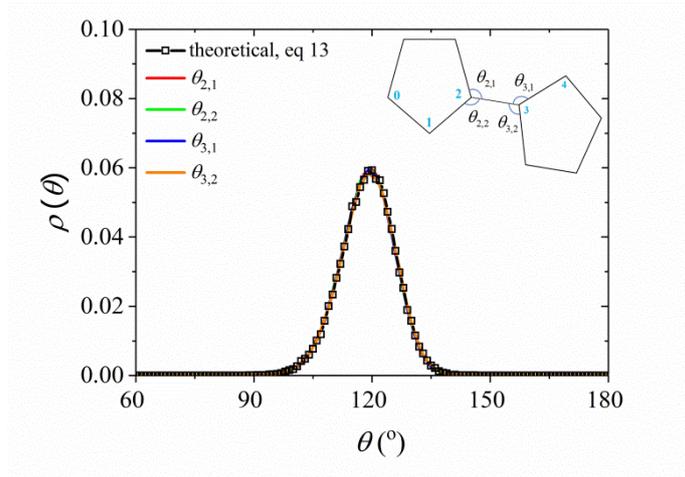

**Figure 9.** Normalized probability distribution of inter-ring bond bending angles $\theta_{2,1}$, $\theta_{2,2}$, $\theta_{3,1}$ and $\theta_{3,2}$ around the bond connecting two successive thiophene rings in an α-6T molecule at $T = 1000$ K. The squares correspond to the theoretical distribution according to eq 13 and the line to the simulation results from our MC algorithm with phantom α-6T chains.

In Appendix B, we also derived an analytical expression for the probability distribution $P(\phi_3)$ of the inter-ring torsion angle $\phi_3$ (or $\phi_{\text{C-S-C-C}}$) using the constraint that all ring atoms are co-planar (see appendix B.3). The result is:

$$\rho(\phi_3) = \frac{\int_0^\pi \sin\theta_{2,1} \exp\left[-\beta U_{\text{bend}}(\theta_{2,1}) - \beta U_{\text{bend}}(\theta_{2,2})\right] d\theta_{2,1}}{\int_0^{2\pi} d\phi_3 \int_0^\pi \sin\theta_{2,1} \exp\left[-\beta U_{\text{bend}}(\theta_{2,1}) - \beta U_{\text{bend}}(\theta_{2,2})\right] d\theta_{2,1}} \quad (14)$$

and is graphichally shown in Figure 10 by the black squares. For comparison, also shown in the same Figure are the simulation results from our MC algorithm with phantom a-6T chains for a mix of moves which guarrantees ergodic simulation. The two sets of data again coincide, which is an another indication that all MC moves have been implemented corectly in our algorithm.

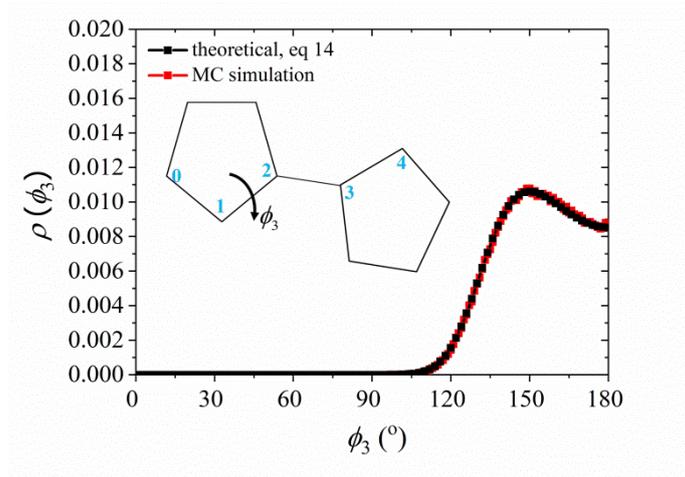

**Figure 10:** Normalized probability distribution of torsion angle $\phi_3$ corresponding to atoms 0, 1, 2 and 3 (also known as the C-S-C-C inter-ring torsion angle) as obtained from the MC simulations with phantom α-6T chains and as computed theoretically (black squares) according to eq 14 ($T = 1000$ K). Because of its symmetry with respect to 180°, only half of the graph is shown.

As far as the torsion angle $\phi_4$ (or $\phi_{\text{S-C-C-S}}$) is concerned (see Figure 1), according to eq 15 this should be uniformly distributed in the interval $[0, 2\pi)$:

$$\rho(\phi_4) = \frac{1}{2\pi} \qquad (15)$$

Graphically, this is shown in Figure 11 together with simulation results from our MC code with phantom α-T6 chains with the following mix of moves: 25 % bias-reptations, 25 % ring-flips, 25 % ConRot's and 25 % end-ring rotations. Our MC simulation results are consistent with the theoretical predictions, which is another indication of the correctness of our MC algorithm.

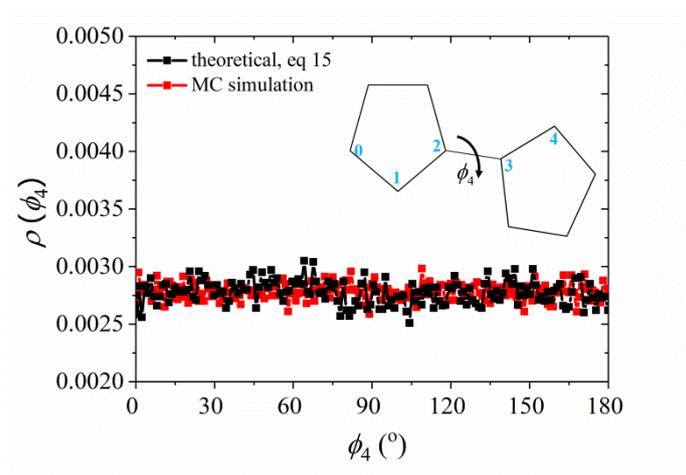

**Figure 11**. Normalized probability distribution of the torsion angle $\phi_4$ corresponding to atoms 1, 2, 3 and 4 (also known as the S-C-C-S torsion angle) as obtained from the MC simulations with phantom α-6T chains (red squares) and as computed theoretically (black squares) according to eq 15 ($T$ = 1000 K). Again, because of its symmetry with respect to 180°, only half of the graph is shown.

**5.2 Computational efficiency.** In addition to the proper implementation of the MC moves, equally important is to examine how efficiently the new algorithm can equilibrate bulk models of oligo- and poly-thiophenes. This can be tested by computing the time needed for the orientational autocorrelation function $\langle \mathbf{u}(t)\cdot\mathbf{u}(0)\rangle$ of the unit vector $\mathbf{u}$ directed along the chain end-to-end vector $\mathbf{R}$ with CPU time ($T$ = 740 K, $P$ = 1 atm) to drop to zero. The vector $\mathbf{R}$ is defined as the vector connecting the CH* (see atom types of Table 1) atomistic-units between the first and last ring along a thiophene molecule. By construction, the rate with which the function $\langle \mathbf{u}(t)\cdot\mathbf{u}(0)\rangle$ drops to zero is a measure of how fast the system de-correlates from its initial configuration; it provides therefore a measure of the efficiency of the simulation method in equilibrating the long-range characteristics of the system. We have used the new algorithm to simulate a bulk system of α-6T molecules at $T$ = 740 K and $P$ = 1 atm, and the result obtained is displayed in Figure 12 where MC iterations have been transformed to CPU time. We chose α-6T because it is a rather well-studied oligo-thiophene both experimentally[23, 24, 76] and computationally.[44] For comparison, we also show in Figure 12 the corresponding curve obtained from simulating the same system not with MC but with the MD method at the same temperature and pressure. To make the comparison fair between the two methods, both MC and MD simulations have been executed serially (i.e., on a single CPU). It turns out that for this small molecule considered here, α-6T, MD is more efficient than MC. However, as the chain length increases, we know that MD will become progressively less and less efficient, because of the fast increase of the longest relaxation time of the molecule with molecular length. We thus expect the new MC algorithm to be more advantageous when longer thiophene molecules are simulated (especially if additional moves based on chain-connectivity altering schemes such as double bridging are introduced[62]) but this is an issue that deserves further investigation.

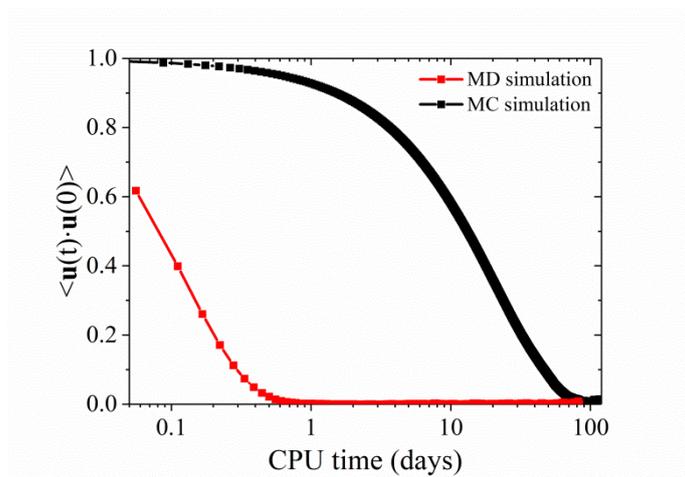

**Figure 12**. Decay of the orientational relaxation function $\langle \mathbf{u}(t) \cdot \mathbf{u}(0) \rangle$ with CPU time. The results refer to a bulk system of 1120 α-6T molecules at $T = 800$ K and $P = 1$ atm. The black curve shows the result obtained from the simulation with the new MC algorithm and the red curve the result obtained from the MD method. CPU times have been corrected so that they correspond to execution using only a single CPU.

**5.3. Density, structural and conformational properties.** As we have already discussed in the Introduction, oligo- and poly-thiophenes exhibit a very rich phase behavior diagram characterized by the appearance of specific liquid crystalline morphologies as they are cooled down to lower temperatures (from their Iso phase at a high enough temperature). As a first step towards such an investigation, we report here simulation predictions from the new algorithm for the volumetric, structural and conformational properties of α-6T.[23, 24, 44] In this paper, we limit our discussion only to the high-temperature behavior of this oligo-thiophene, above its Iso-to-Nem phase transition; and we will compare our results directly with the corresponding MD simulation predictions based on an all-atom model. A detailed analysis of the liquid-crystalline phase behavior of this thiophene-based oligomer from simulations with the new algorithm at lower temperatures will be presented in a future contribution.

At temperatures above approximately 700 K, α-6T is in its isotropic molten state, exhibiting no long range orientational or translational ordering. Important properties to check at these temperatures include the density, the distribution of inter-ring torsion angles, and several radial pair correlation functions. We have obtained predictions of all these properties from the new algorithm at $T = 740$ K (almost 50 K - 60 K above the first transition point from the Iso-to-Nem phase[44] and we have compared them with the corresponding sets of data from additional MD simulations using the *flexible* united-atom model.

Table 5 shows the predicted densities the average square end-to-end distance $\langle R^2 \rangle$ of α-6T molecules from the two simulations and it is evident that there is an excellent agreement between the two methods.

**Table 5.** MC and MD simulation predictions for the density, average-square end-to-end distance $\langle R^2 \rangle$ and percentage of t-t-t-t-t, c-t-t-t-t, c-c-t-t-t, c-c-c-t-t, c-c-c-c-t and c-c-c-c-c conformational states in α-6T ($T = 740$ K, $P = 1$ atm). The last three columns indicate the percentage of c-t-t-t-t states that contain one c-state either in the outer or in the inner inter-ring bonds.

| method | density (g cm$^{-3}$) | $\langle R^2 \rangle$ (Å$^2$) | t-t-t-t (%) | c-t-t-t (%) | c-c-t-t (%) | c-c-c-t (%) | c-c-c-t (%) | c-c-c-c (%) | c-t-t-t (%) | t-c-t-t (%) | t-t-c-t-t (%) |
|---|---|---|---|---|---|---|---|---|---|---|---|
| MC | 0.97 ± 0.01 | 418.0 ±1.4 | 22.3 ±1.6 | 40.2 ±1.9 | 27.5 ±1.8 | 8.7 ±1.2 | 1.2 ±0.5 | 0.1 ±0.1 | 42.5 ±2.9 | 36.9 ±3.0 | 20.6 ±2.4 |
| MD | 0.97 ± 0.01 | 418.0 ±1.4 | 16.1 ±1.1 | 35.5 ±1.7 | 31.6 ±1.4 | 13.7 ±1.0 | 2.9 ±0.6 | 0.2 ±0.2 | 43.6 ±3.4 | 37.3 ±3.5 | 19.1 ±2.9 |

The isotropic character of the system at $T = 740$ K is verified by computing several radial pair correlation functions. These include the radial pair correlation function $g_{\text{CoM}}(r)$ of the centers-of-mass (CoM) of α-6T molecules (see Figure 13), and the intermolecular radial pair correlation functions $g_{\text{inter}}^{\text{S-S}}(r)$ and $g_{\text{inter}}^{\text{C-C}}(r)$ of S-S (see Figure 14a) and C-C pairs (see Figure 14b), respectively. Figure 13 shows that a structureless $g_{\text{CoM}}(r)$-vs.-$r$ plot is obtained from both the MC and MD simulations indicative of complete lack of long-range translational order, consistent with the fact that at this high temperature α-6T is in its amorphous molten state. The first peak of the correlation function is quite diffuse and can be attributed to three contributions between nearest-neighbor inter-chain rings. Similar conclusions are drawn by observing the $g_{\text{inter}}^{\text{S-S}}(r)$-vs.-$r$ and $g_{\text{inter}}^{\text{C-C}}(r)$-vs.-$r$ plots in Figures 14a and 14b, respectively. The graphs from the two different simulation methods are again structureless revealing complete lack of long-range positional order at long intermolecular distances.

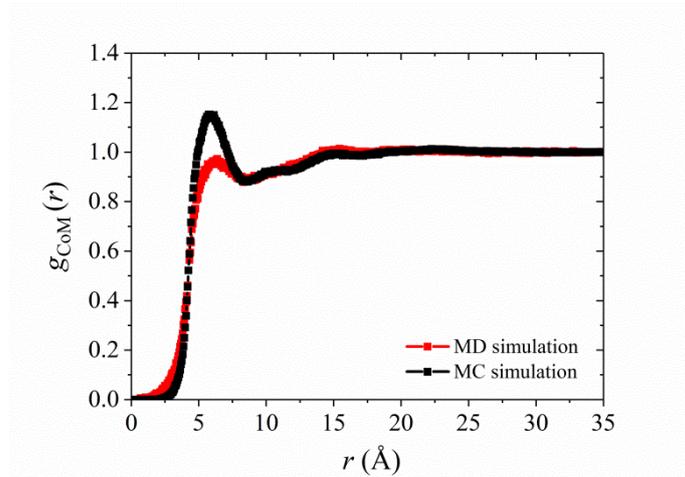

**Figure 13.** MC and MD simulation predictions for the radial pair correlation function $g_{CoM}(r)$ of the centers-of-mass of α-6T molecules at $T = 740$ K and $P = 1$ atm.

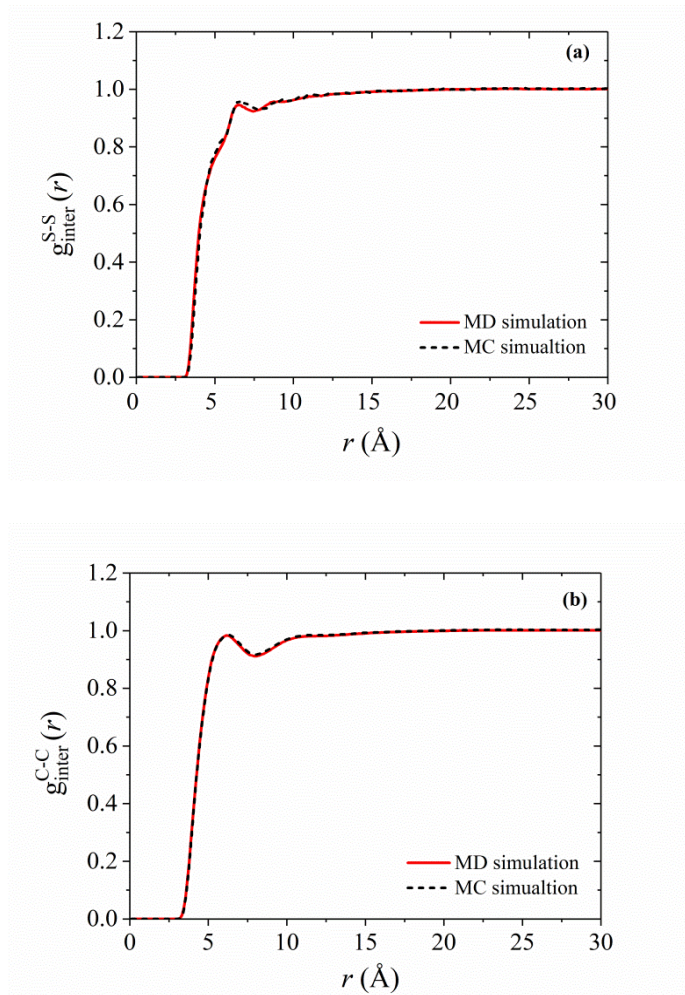

**Figure 14**. MC and MD simulation predictions for the intermolecular radial pair correlation functions $g_{inter}^{S\text{-}S}(r)$ and $g_{inter}^{C\text{-}C}(r)$ of S-S (a) and C-C pairs (b) of α-6T molecules at $T = 740$ K and $P = 1$ atm.

It is clear that the conformation of an α-6T chain is essentially defined by the distribution of the inter-ring torsion angles. For two successive rings, the S-C-C-S torsion angle defines their degree of co-planarity while the C-S-C-C torsion angle their bending. The distributions of these angles at $T = 740$ K are illustrated in Figure 15. For each torsion angle, the distribution calculated for the two models are of the same form. The S-C-C-S distribution (see Figure 15a) shows that successive rings tend to be co-planar since two intense peaks are located at $\phi = 0°$ (*cis* state) and $\phi = 180°$ (*trans* state). Moreover, the C-S-C-C distribution (see Figure 15b) presents one intense peak at $\phi = 180°$ indicating limited bending between successive rings. This peak is sharper in the curve computed from the MC simulations implying that bending is less favorable compared to that from the MD simulations. This difference between the two sets of simulation data is characteristic of the stiffer character of the united-atom model employed in the MC simulations (e.g., constant bond lengths and fully planar geometry of the thiophene ring).

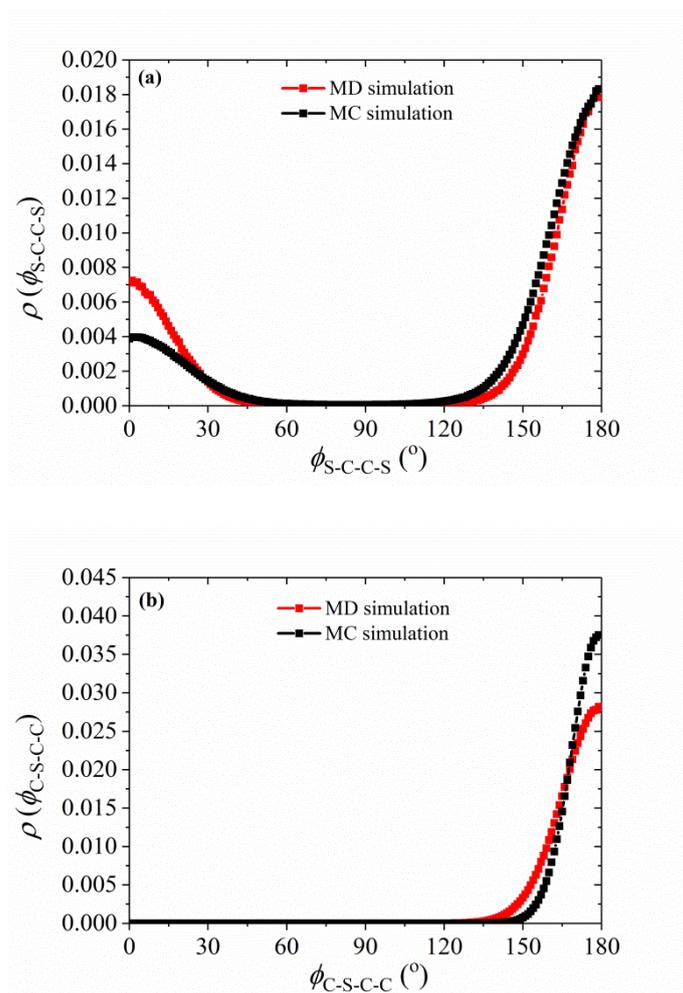

**Figure 15.** Distribution of the S-C-C-S -also names as $\phi_4$- (a) and C-S-C-C -also names as $\phi_3$- (b) inter-ring torsion angles from the MC and MD simulations, at $T = 740$ K and $P = 1$ atm. Only the half of the horizontal axis is shown due to the symmetry of the curves.

Similar conclusions are drawn by observing the $w_{intra}^{S-S}(r)$-vs.-$r$ and $w_{intra}^{C-C}(r)$-vs.-$r$ plots of the S-S and C-C intra-chain pair density functions in Figures 16a and 16b, respectively. The graphs from the two different simulation methods are very similar, differing only in the exact locations of the various peaks; the latter correspond to characteristic S-S or C-C distances along an α-6T molecule between sulfur or carbon pairs located one, two, three, etc. thiophene rings apart. The first two peaks, in particular, are found at distances approximately equal to 3.3 Å and 4.5 Å from the MD simulations and equal to 3.2 Å and 4.4 Å from the MC simulations, which are directly related to the *cis* $(\phi = 0°)$ and *trans* $(\phi = \pm 180°)$ conformations of the inter-ring S-C-C-S torsion angle, respectively. Essentially, the value of the S-C-C-S torsion angle $\phi_{S-C-C-S}$ defines the relevant arrangement of two successive thiophene units: if -$\phi_{S-C-C-S} \in [-90°, 90°]$ the rings are in the *syn-SS-parallel* configuration denoted as the 'c-' state whereas if $\phi_{S-C-C-S} \in [-180°, 90°] \cup [90°, 180°]$ the rings are in the *anti-SS-parallel* configuration denoted as the 't-' state. The comparison of the heights of the peaks indicates that *anti-SS-parallel* configurations are favored in the isotropic state. The first three intense peaks in the $w_{intra}^{C-C}(r)$-vs.-$r$ plot correspond to nearest-neighbor carbon sites (see Figure 16b). In all cases described here, the intra-atomic distances for the *stiff* model implemented in the MC simulations are slightly smaller than the corresponding ones obtained from the MD simulations using the *flexible* model.

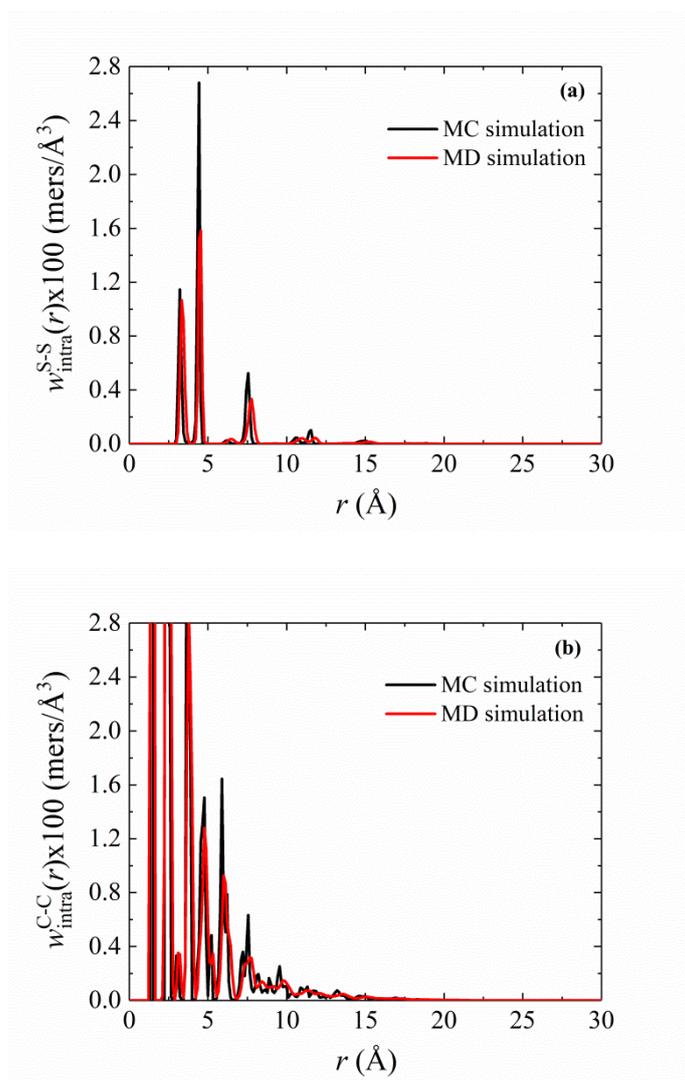

**Figure 16.** MC and MD simulation predictions for the intra-chain pair density functions $w_{intra}^{S-S}(r)$ and $w_{intra}^{C-C}(r)$ of: (a) S-S pairs, and (b) C-C pairs of α-6T molecules at $T = 740$ K and $P = 1$ atm.

The chain conformations can be grouped by counting the number of t- and c- states of the inter-ring angles (or the successive rings configurations) independently of their position in the chain. Hence, in the case of α-6T molecule five conformational states are defined and denoted by a sequence of five letters, i.e., t-t-t-t-t, c-t-t-t-t, c-c-t-t-t, c-c-c-t-t, c-c-c-c-t and c-c-c-c-c. Each group consists of a number of conformations that occur by mutual permutation of the constituent c- and t-states. Typical t-t-t-t-t and c-t-t-t-t states are depicted schematically in Figures 17a and 17b, respectively. The percentages of these conformations of the chains as computed as ensemble averages from the MD and MC simulations for various temperatures are given in Table 3. It is evident that the more c-states are included in a conformational group, the less probable the appearance of this group is.

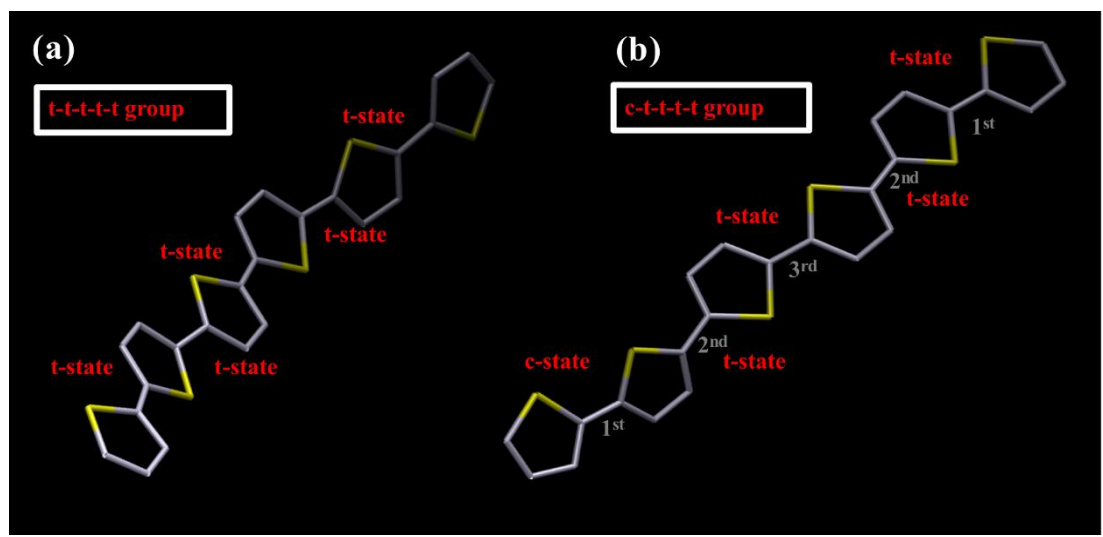

**Figure 17.** Typical atomistic configurations of an α-6T molecule with the following sequence of conformational states: (a) t-t-t-t-t and (b) c-t-t-t-t. Inter-ring bonds in (b) are denoted according to their local numbers ($1^{st}$, $2^{nd}$, $3^{rd}$). Sulfur atoms are shown in yellow and carbon atoms in grey.

In the Iso phase the dominant conformational states are the c-t-t-t-t and c-c-t-t-t rather than the all-*trans* (t-t-t-t-t). Nevertheless, one can notice that the total percentage of t-states that are included in the conformations of the chains is approximately 68 % (see Table 3). For example, the percentage of t-states in the system at 740 K which is calculated as $\sum_{i \in \text{groups}} (\text{group\%}) \times (\# \, t \, \text{states})_i$ comes out to be ≈ 69.6 % from the MD simulation and ≈ 74.6 % from the MC simulation. This finding is also consistent with the distribution of the torsion angles in Figure 15. It is also the reason why the intensity of the peak that corresponds to t- states is higher than the one that corresponds to c-states regarding the first group of 'twin' peaks that is shown in Figure 16a.

Both models predict that in the isotropic phase at $T = 740$ K the population of c-t-t-t-t conformations is the highest (between 35% and 41% depending on the model) while that of c-c-c-c-c the lowest. Furthermore, chain conformations that contain at least two c states are very high (between 28% and 32% depending on the model) at this temperature ($T$ = 740 K). For further details, see Table 5.

Given the high percentage of c-t-t-t-t conformation at $T$ = 740 K, we have checked where the c-state is located along the α-6T molecule. The results are displayed in the last three columns of Table 5 from which we can observe that the largest percentage of c-states are located at the outer inter-ring bond (the 1st outer inter-ring bond in Figure 17). Slightly smaller is their percentage in the 2nd inter-ring bond, whereas the least percentage is found in the 3rd (and inner-most) inter-ring bond (see Figure 17).

**6. Conclusions**

We have implemented state-of-the-art MC moves, originally proposed for considerably simpler polymer structures and architectures such as linear and branched polyethylene, for the atomistic simulation of thiophene oligomers and polymers in the united atom representation. In our implementation of all moves, thiophene ring atoms are assumed to remain rigid and strictly co-planar; inter-ring torsion and bond bending angles, on the other hand, are allowed to fluctuate subject to suitable potential energy functions. Given the complexity of the moves for the molecules dealt with here, we ran several test simulations with the new algorithm using phantom chains (for which we could obtain analytical expressions for the distribution of bond bending and torsion angles) to confirm that all of them were correctly implemented in our code. For all moves, the phantom chain tests were successful.

Test simulations with the new MC algorithm of an important thiophene oligomer for organic electronic applications, α-sexithiophene (α-6T), at high enough temperatures (above its Nem-to-Iso phase transition) provided predictions for its thermodynamic, structural and conformational properties that are in remarkable agreement with additional simulation data obtained from a more detailed, fully *flexible* model.

This work focused entirely on the high temperature behavior of α-6T where it remains isotropic. In a future publication, we will use the new algorithm to simulate α-unsubstituted OTs at lower temperatures to examine the appearance of liquid crystalline phases. We expect the new algorithm to be particularly promising for exploring the phase behavior of more complex polythiophene molecules which cannot be addressed by brute force MD due to the extremely long relaxation times characterizing chain dynamics in high molecular weight polymers.

**Acknowledgment**. Financial support for this study has been provided by the General Secretariat for Research and Technology (GSRT) and the Hellenic Foundation for Research and Innovation (HFRI) under the scholarship announcement: 'First HFRI call for the financial support of doctoral candidates', code 1460. Financial support has also been provided through the National Project titled: General method for the simulation of self-organization in nanostructured polymeric systems (GENESIS, code 1010 – D655) which is part of the Action "ARISTEIA - Education and Lifelong Learning" Programme and is supported both by the European Union (European Social Fund-ESF) and national funds. We are grateful to the Greek Research & Technology Network (GRNET) for generous allocation of computational time on the National HPC facility-ARIS, located in Athens, Greece, under project THIOSIM (pr005043). We thank Dr. Dimitris Dellis from GRNET for his support on several technical aspects of the present work.

**References**

1.  Fichou, D. Handbook of Oligo- and Polythiophenes. *Wiley-VCH* **1999,** Weinheim, Germany.

2.  Roncali, J. Conjugated Poly(Thiophenes) - Synthesis, Functionalization, and Applications. *Chem. Rev.* **1992,** 92 (4), 711-738 DOI: 10.1021/Cr00012a009.

3. Fichou, D. Structural order in conjugated oligothiophenes and its implications on opto-electronic devices. *J. Mater. Chem*. **2000,** 10 (3), 571-588 DOI: 10.1039/A908312j.

4. Mena-Osteritz, E. Superstructures of self-organizing thiophenes. *Adv. Mater.* **2002,** 14 (8), 609-616 DOI: 10.1002/1521-4095(20020418)14:8<609::Aid-Adma609>3.0.Co;2-7.

5. Horowitz, G.; Hajlaoui, M. E. Mobility in polycrystalline oligothiophene field-effect transistors dependent on grain size. *Adv. Mater.* **2000,** 12 (14), 1046-1050 DOI: 10.1002/1521-4095(200007)12:14<1046::Aid-Adma1046>3.0.Co;2-W.

6. Dimitrakopoulos, C. D.; Malenfant, P. R. L. Organic thin film transistors for large area electronics. *Adv. Mater.* **2002,** 14 (2), 99-117 DOI: 10.1002/1521-4095(20020116)14:2<99::Aid-Adma99>3.0.Co;2-9.

7. Katz, H. E.; Bao, Z. The physical chemistry of organic field-effect transistors. *J. Phys. Chem. B* **2000,** 104 (4), 671-678 DOI: 10.1021/Jp992853n.

8. Horowitz, G. Field-effect transistors based on short organic molecules. *J. Mater. Chem.* **1999,** 9 (9), 2021-2026 DOI: 10.1039/A902242b.

9. Horowitz, G. Organic field-effect transistors. *Adv. Mater.* **1998,** 10 (5), 365-377 DOI: 10.1002/(Sici)1521-4095(199803)10:5<365::Aid-Adma365>3.0.Co;2-U.

10. Lovinger, A. J.; Rothberg, L. J. Electrically active organic and polymeric materials for thin-film-transistor technologies. *J. Mater. Res.* **1996,** 11 (6), 1581-1592 DOI: 10.1557/Jmr.1996.0198.

11. Nalwa, H. S., *Handbook of Advanced Electronic and Photonic Materials and Devices*. 2000.

12. Burroughes, J. H.; Bradley, D. D. C.; Brown, A. R.; Marks, R. N.; Mackay, K.; Friend, R. H.; Burn, P. L.; Holmes, A. B. Light-Emitting-Diodes Based on Conjugated Polymers. *Nature* **1990,** 347 (6293), 539-541 DOI: 10.1038/347539a0.

13. Spanggaard, H.; Krebs, F. C. A brief history of the development of organic and polymeric photovoltaics. *Sol. Energ. Mat. Sol. C* **2004,** 83 (2-3), 125-146 DOI: 10.1016/j.solmat.2004.02.021.

14. Smith, A. P.; Smith, R. R.; Taylor, B. E.; Durstock, M. F. An investigation of poly(thienylene vinylene) in organic photovoltaic device. *Chem. Mater.* **2004,** 16 (23), 4687-4692 DOI: 10.1021/cm049447n.

15. Hedley, G. J.; Ward, A. J.; Alekseev, A.; Howells, C. T.; Martins, E. R.; Serrano, L. A.; Cooke, G.; Ruseckas, A.; Samuel, I. D. W. Determining the optimum morphology in high-performance polymer-fullerene organic photovoltaic cells. *Nat. Commun.* **2013,** 4, DOI: Artn 2867 10.1038/Ncomms3867.

16. Olivier, Y.; Niedzialek, D.; Lemaur, V.; Pisula, W.; Mullen, K.; Koldemir, U.; Reynolds, J. R.; Lazzaroni, R.; Cornil, J.; Beljonne, D. 25th Anniversary Article: High-Mobility Hole and Electron Transport Conjugated Polymers: How Structure Defines Function. *Adv. Mater.* **2014,** 26 (14), 2119-2136 DOI: 10.1002/adma.201305809.

17. Garnier, F. Organic-based electronics a la carte. *Accounts Chem. Res.* **1999,** 32 (3), 209-215 DOI: 10.1021/Ar9800340.

18. Curtis, M. D.; Cao, J.; Kampf, J. W. Solid-state packing of conjugated oligomers: from pi-stacks to the herringbone structure. *J. Am. Chem. Soc.* **2004,** 126 (13), 4318-4328 DOI: 10.1021/ja0397916.

19. Facchetti, A.; Yoon, M. H.; Stern, C. L.; Hutchison, G. R.; Ratner, M. A.; Marks, T. J. Building blocks for N-type molecular and polymeric electronics. Perfluoroalkyl- versus alkyl-functionalized ligothiophenes (nTs; n=2-6). Systematic synthesis, spectroscopy, electrochemistry, and solid-state organization. *J. Am. Chem. Soc.* **2004,** 126 (41), 13480-13501 DOI: 10.1021/ja048988a.

20. Mishra, A.; Ma, C. Q.; Bauerle, P. Functional Oligothiophenes: Molecular Design for Multidimensional Nanoarchitectures and Their Applications. *Chem. Rev.* **2009,** 109 (3), 1141-1276 DOI: 10.1021/cr8004229.

21. Kuiper, S.; Jager, W. F.; Dingemans, T. J.; Picken, S. J. Liquid crystalline properties of all symmetric p-phenylene and 2,5-thiophene pentamers. *Liq. Cryst.* **2009,** 36 (4), 389-396 DOI: 10.1080/02678290902923410.

22. Kuiper, S. Synthesis, Mesophase Behaviour and Charge Mobility of Phenyl-Thiophene Pentamers. 2011.

23. Taliani, C.; Zamboni, R.; Ruani, G.; Rossini, S.; Lazzaroni, R. New Rigid Rod Liquid-Crystal Molecule Precursor of Conjugated Polymers - Alpha-Sexithienyl. *J. Mol. Electron.* **1990,** 6 (4), 225-226.

24. Destri, S.; Mascherpa, M.; Porzio, W. Mesophase Formation in Alpha-Sexithienyl at High-Temperature - an X-Ray-Diffraction Study. *Adv. Mater.* **1993,** 5 (1), 43-45 DOI: 10.1002/adma.19930050108.

25. Porzio, W.; Destri, S.; Mascherpa, M.; Bruckner, S. Structural Aspects of Oligothienyl Series from X-Ray-Powder Diffraction Data. *Acta Polym.* **1993,** 44 (6), 266-272 DOI: 10.1002/actp.1993.010440602.

26. Kauffmann, T. From the Principle of Areno-Analogy to Heterocyclopolyaromatic Compounds. *Angew. Chem.* **1979,** 18, 1-90 DOI: 10.1002/anie.197900013.

27. Fichou, D.; Teulade-Fichou, M. P.; Horowitz, G.; Demanze, F. Thermal and optical characterization of high purity alpha-octithiophene. *Adv. Mater.* **1997,** 9 (1), 75-79 DOI: 10.1002/adma.19970090118.

28. Allinger, N.; Yuh, Y. Quantum chemistry program exchange. *Indiana University, Program* **1980,** 395.

29. Allinger, N. L.; Kok, R. A.; Imam, M. R. Hydrogen-Bonding in Mm2. *J. Comput. Chem.* **1988,** 9 (6), 591-595 DOI: 10.1002/jcc.540090602.

30. Ferro, D. R.; Porzio, W.; Destri, S.; Ragazzi, M.; Bruckner, S. Application of molecular mechanics to refine and understand the crystal structure of polythiophene and its oligomers. *Macromol. Theor. Simul.* **1997,** 6 (4), 713-727 DOI: 10.1002/mats.1997.040060403.

31. Kobayashi, M.; Chen, J.; Chung, T.-C.; Moraes, F.; Heeger, A.; Wudl, F. Synthesis and properties of chemically coupled poly (thiophene). *Synthetic Met.* **1984,** 9 (1), 77-86 DOI: 10.1016/0379-6779(84)90044-4.

32. Raos, G.; Famulari, A.; Marcon, V. Computational reinvestigation of the bithiophene torsion potential. *Chem. Phys. Lett.* **2003,** 379 (3-4), 364-372 DOI: 10.1016/j.cplett.2003.08.060.

33. Marcon, V.; Raos, G. Molecular modeling of crystalline oligothiophenes: Testing and development of improved force fields. *J. Phys. Chem. B* **2004,** 108 (46), 18053-18064 DOI: 10.1021/jp047128d.

34. Allinger, N. L.; Yuh, Y. H.; Lii, J. H. Molecular Mechanics - the Mm3 Force-Field for Hydrocarbons .1. *J. Am. Chem. Soc.* **1989,** 111 (23), 8551-8566 DOI: 10.1021/Ja00205a001.

35. Lii, J. H.; Allinger, N. L. Molecular Mechanics - the Mm3 Force-Field for Hydrocarbons .3. The Vanderwaals Potentials and Crystal Data for Aliphatic and Aromatic-Hydrocarbons. *J. Am. Chem. Soc.* **1989,** 111 (23), 8576-8582 DOI: 10.1021/Ja00205a003.

36. Marcon, V.; Raos, G. Free energies of molecular crystal surfaces by computer simulation: Application to tetrathiophene. *J. Am. Chem. Soc.* **2006,** 128 (5), 1408-1409 DOI: 10.1021/ja056548t.

37. Moreno, M.; Casalegno, M.; Raos, G.; Meille, S. V.; Po, R. Molecular Modeling of Crystalline Alkylthiophene Oligomers and Polymers. *J. Phys. Chem. B* **2010,** 114 (4), 1591-1602 DOI: 10.1021/jp9106124.

38. Jorgensen, W. L.; Maxwell, D. S.; TiradoRives, J. Development and testing of the OPLS all-atom force field on conformational energetics and properties of organic liquids. *J. Am. Chem. Soc.* **1996,** 118 (45), 11225-11236 DOI: 10.1021/Ja9621760.

39. Jorgensen, W. L.; Tirado-Rives, J. Potential energy functions for atomic-level simulations of water and organic and biomolecular systems. *P. Natl. Acad. Sci. USA* **2005,** 102 (19), 6665-6670 DOI: 10.1073/pnas.0408037102.

40. Darling, S. B.; Sternberg, M. Importance of Side Chains and Backbone Length in Defect Modeling of Poly(3-alkylthiophenes). *J. Phys. Chem. B* **2009,** 113 (18), 6215-6218 DOI: 10.1021/jp808045j.

41. Huang, D. M.; Faller, R.; Do, K.; Moule, A. J. Coarse-Grained Computer Simulations of Polymer/Fullerene Bulk Heterojunctions for Organic Photovoltaic Applications. *J. Chem. Theory Comput.* **2010,** 6 (2), 526-537 DOI: 10.1021/ct900496t.

42. Schwarz, K. N.; Kee, T. W.; Huang, D. M. Coarse-grained simulations of the solution-phase self-assembly of poly(3-hexylthiophene) nanostructures. *Nanoscale* **2013,** 5 (5), 2017-2027 DOI: 10.1039/c3nr33324h.

43. Alberga, D.; Mangiatordi, G. F.; Torsi, L.; Lattanzi, G. Effects of Annealing and Residual Solvents on Amorphous P3HT and PBTTT Films. *J. Phys. Chem. C* **2014,** 118 (16), 8641-8655 DOI: 10.1021/jp410936t.

44. Pizzirusso, A.; Savini, M.; Muccioli, L.; Zannoni, C. An atomistic simulation of the liquid-crystalline phases of sexithiophene. *J. Mater. Chem.* **2011,** 21 (1), 125-133 DOI: 10.1039/c0jm01284j.

45. Weiner, S. J.; Kollman, P. A.; Case, D. A.; Singh, U. C.; Ghio, C.; Alagona, G.; Profeta, S.; Weiner, P. A New Force-Field for Molecular Mechanical Simulation of Nucleic-Acids and Proteins. *J. Am. Chem. Soc.* **1984,** 106 (3), 765-784 DOI: 10.1021/Ja00315a051.

46. Cornell, W. D.; Cieplak, P.; Bayly, C. I.; Gould, I. R.; Merz, K. M.; Ferguson, D. M.; Spellmeyer, D. C.; Fox, T.; Caldwell, J. W.; Kollman, P. A. A 2nd Generation Force-Field for the Simulation of Proteins, Nucleic-Acids, and Organic-Molecules. *J. Am. Chem. Soc.* **1995,** 117 (19), 5179-5197 DOI: 10.1021/Ja00124a002.

47. Zhang, G. L.; Pei, Y.; Ma, J.; Yin, K. L.; Chen, C. L. Packing structures and packing effects on excitation energies of amorphous phase oligothiophenes. *J. Phys. Chem. B* **2004,** 108 (22), 6988-6995 DOI: 10.1021/Jp037882j.

48. Sun, H. Ab-Initio Calculations and Force-Field Development for Computer-Simulation of Polysilanes. *Macromolecules* **1995,** 28 (3), 701-712 DOI: 10.1021/Ma00107a006.

49. Sun, H.; Mumby, S. J.; Maple, J. R.; Hagler, A. T. Ab-Initio Calculations on Small-Molecule Analogs of Polycarbonates. *J. Phy.s Chem.* **1995,** 99 (16), 5873-5882 DOI: 10.1021/J100016a022.

50. Hill, J. R.; Sauer, J. Molecular Mechanics Potential for Silica and Zeolite Catalysts Based on Ab-Initio Calculations .1. Dense and Microporous Silica. *J. Phys. Chem.* **1994,** 98 (4), 1238-1244 DOI: 10.1021/J100055a032.

51. Hwang, M. J.; Stockfisch, T. P.; Hagler, A. T. Derivation of Class-Ii Force-Fields .2. Derivation and Characterization of a Class-Ii Force-Field, Cff93, for the Alkyl Functional-Group and Alkane Molecules. *J. Am. Chem. Soc.* **1994,** 116 (6), 2515-2525 DOI: 10.1021/Ja00085a036.

52. Curco, D.; Aleman, C. Computational tool to model the packing of polycyclic chains: Structural analysis of amorphous polythiophene. *J. Comput. Chem.* **2007,** 28 (10), 1743-1749 DOI: 10.1002/jcc.20687.

53. Vukmirovic, N.; Wang, L. W. Electronic Structure of Disordered Conjugated Polymers: Polythiophenes. *J. Phys. Chem. B* **2009,** 113 (2), 409-415 DOI: 10.1021/jp808360y.

54. Maple, J. R.; Hwang, M. J.; Stockfisch, T. P.; Dinur, U.; Waldman, M.; Ewig, C. S.; Hagler, A. T. Derivation of Class-Ii Force-Fields .1. Methodology and Quantum Force-Field for the Alkyl Functional-Group and Alkane Molecules. *J. Comput. Chem.* **1994,** 15 (2), 162-182 DOI: 10.1002/jcc.540150207.

55. Theodorou, D. N. Progress and Outlook in Monte Carlo Simulations. *Ind. Eng. Chem. Res.* **2010,** 49 (7), 3047-3058 DOI: 10.1021/ie9019006.

56. Lee, C. K.; Pao, C. W. Nanomorphology Evolution of P3HT/PCBM Blends during Solution-Processing from Coarse-Grained Molecular Simulations. *J. Phys. Chem. C* **2014,** 118 (21), 11224-11233 DOI: 10.1021/jp501323p.

57. de Pablo, J. J.; Curtin, W. A. Multiscale modeling in advanced materials research: Challenges, novel methods, and emerging applications. *Mrs Bull* **2007,** 32 (11), 905-911 DOI: 10.1557/Mrs2007.187.

58. Ottinger, H. C. Systemic coarse graining: "Four lessons and a caveat" from nonequilibrium statistical mechanics. *Mrs Bull* **2007,** 32 (11), 936-940 DOI: Doi 10.1557/Mrs2007.191.

59. Allen, M. P.; Tildesley, D. J., *Computer simulation of liquids*. Oxford university press: 2017.

60. Frenkel, D.; Smit, B., *Understanding molecular simulation: from algorithms to applications*. Elsevier: 2001; Vol. 1.

61. Metropolis, N.; Rosenbluth, A. W.; Rosenbluth, M. N.; Teller, A. H.; Teller, E. Equation of state calculations by fast computing machines. *J. Chem. Phys.* **1953,** 21 (6), 1087-1092. DOI: 10.1063/1.1699114.

62. Karayiannis, N. C.; Giannousaki, A. E.; Mavrantzas, V. G.; Theodorou, D. N. Atomistic Monte Carlo simulation of strictly monodisperse long polyethylene melts through a generalized chain bridging algorithm. *J. Chem. Phys.* **2002,** 117 (11), 5465-5479 DOI: 10.1063/1.1499480.

63. Karayiannis, N. C.; Mavrantzas, V. G.; Theodorou, D. N. A novel Monte Carlo scheme for the rapid equilibration of atomistic model polymer systems of precisely defined molecular architecture. *Phys. Rev. Lett.* **2002,** 88 (10), DOI: Artn 10550310.1103/Physrevlett.88.105503.

64. Daoulas, K. C.; Terzis, A. F.; Mavrantzas, V. G. Detailed atomistic Monte Carlo simulation of grafted polymer melts. I. Thermodynamic and conformational properties. *J. Chem. Phys.* **2002,** 116 (24), 11028-11038 DOI: 10.1063/1.1478055.

65. Doxastakis, M.; Mavrantzas, V. G.; Theodorou, D. N. Atomistic Monte Carlo simulation of cis-1,4 polyisoprene melts. I. Single temperature end-bridging Monte Carlo simulations. *J. Chem. Phys.* **2001,** 115 (24), 11339-11351 DOI: Doi 10.1063/1.1416490.

66. Mavrantzas, V. G.; Boone, T. D.; Zervopoulou, E.; Theodorou, D. N. End-bridging Monte Carlo: A fast algorithm for atomistic simulation of condensed phases of long polymer chains. *Macromolecules* **1999,** 32 (15), 5072-5096 DOI: 10.1021/Ma981745g.

67. Zervopoulou, E.; Mavrantzas, V. G.; Theodorou, D. N. A new Monte Carlo simulation approach for the prediction of sorption equilibria of oligomers in polymer melts: Solubility of long alkanes in linear polyethylene. *J. Chem. Phys.* **2001,** 115 (6), 2860-2875 DOI: 10.1063/1.1383050.

68. Alexiadis, O.; Harmandaris, V. A.; Mavrantzas, V. G.; Delle Site, L. Atomistic simulation of alkanethiol self-assembled monolayers on different metal surfaces via a quantum, first-principles parametrization of the sulfur-metal interaction. *J. Phys. Chem. C* **2007,** 111 (17), 6380-6391 DOI: 10.1021/jp067347u.

69. Alexiadis, O.; Daoulas, K. C.; Mavrantzas, V. G. An efficient Monte Carlo algorithm for the fast equilibration and atomistic simulation of alkanethiol self-assembled monolayers on a Au(111) substrate. *J. Phys. Chem. B* **2008,** 112 (4), 1198-1211 DOI: 10.1021/jp076417+.

70. Peristeras, L. D.; Economou, I. G.; Theodorou, D. N. Structure and volumetric properties of linear and triarm star polyethylenes from atomistic Monte Carlo simulation using new internal rearrangement moves. *Macromolecules* **2005,** 38 (2), 386-397 DOI: 10.1021/ma048364p.

71. Facchetti, A. π-Conjugated polymers for organic electronics and photovoltaic cell applications. *Chem. Mater.* **2010,** 23 (3), 733-758 DOI: 10.1002/cphc.200900652.

72. Tiberio, G.; Muccioli, L.; Berardi, R.; Zannoni, C. How Does the Trans-Cis Photoisomerization of Azobenzene Take Place in Organic Solvents? *Chemphyschem.* **2010,** 11 (5), 1018-1028 DOI: 10.1002/cphc.200900652.

73. Wildman, J.; Repiscak, P.; Paterson, M. J.; Galbraith, I. General Force-Field Parametrization Scheme for Molecular Dynamics Simulations of Conjugated Materials in Solution. *J. Chem. Theory Comput.* **2016,** 12 (8), 3813-3824 DOI: 10.1021/acs.jctc.5b01195.

74. Mayo, S. L.; Olafson, B. D.; Goddard, W. A. Dreiding - a Generic Force-Field for Molecular Simulations. *J. Phys. Chem.* **1990,** 94 (26), 8897-8909 DOI: 10.1021/J100389a010.

75. Bhatta, R. S.; Yimer, Y. Y.; Tsige, M.; Perry, D. S. Conformations and torsional potentials of poly(3-hexylthiophene) oligomers: Density functional calculations up to the dodecamer. *Comput. Theor. Chem.* **2012,** 995, 36-42 DOI: 10.1016/j.comptc.2012.06.026.

76. Melucci, M.; Barbarella, G.; Zambianchi, M.; Di Pietro, P.; Bongini, A. Solution-phase microwave-assisted synthesis of unsubstituted and modified alpha-quinque- and sexithiophenes. *J. Org. Chem.* **2004,** 69 (14), 4821-4828 DOI: 10.1021/jo035723q.

77. Ramos, J.; Peristeras, L. D.; Theodorou, D. N. Monte Carlo simulation of short chain branched polyolefins in the molten state. *Macromolecules* **2007,** 40 (26), 9640-9650 DOI: 10.1021/ma071615k.

78. Scienomics, M. P., Version 3.4.2, France. 2015. Available online: http://www.scienomics.com/ (accessed on 30 September 2016)

79. Plimpton, S. Fast Parallel Algorithms for Short-Range Molecular-Dynamics. *J. Comput. Phys.* **1995,** 117 (1), 1-19 DOI: 10.1006/jcph.1995.1039.

80. Nose, S. Constant Temperature Molecular-Dynamics Methods. *Prog. Theor. Phys. Supp.* **1991,** (103), 1-46 DOI: 10.1143/PTPS.103.1.

81. Hoover, W. G. Constant-Pressure Equations of Motion. *Phys. Rev. A* **1986,** 34 (3), 2499-2500 DOI: 10.1103/PhysRevA.34.2499.

82. Doxastakis, M.; Mavrantzas, V. G.; Theodorou, D. N. Atomistic Monte Carlo simulation of cis-1,4 polyisoprene melts. II. Parallel tempering end-bridging Monte Carlo simulations. *J. Chem. Phys.* **2001,** 115 (24), 11352-11361 DOI: 10.1063/1.1416491.

83. Karayiannis, N. C.; Giannousaki, A. E.; Mavrantzas, V. G. An advanced Monte Carlo method for the equilibration of model long-chain branched polymers with a well-defined molecular architecture: Detailed atomistic simulation of an H-shaped polyethylene melt. *J. Chem. Phys.* **2003,** 118 (6), 2451-2454 DOI: 10.1063/1.1543580.

84. Tsourtou, F. D.; Alexiadis, O.; Mavrantzas, V. G.; Kolonias, V.; Housos, E. Atomistic Monte Carlo and molecular dynamics simulation of the bulk phase self-assembly of semifluorinated alkanes. *Chem. Eng. Sci.* **2015,** 121, 32-50 DOI: 10.1016/j.ces.2014.09.009.

85. Vacatello, M.; Avitabile, G.; Corradini, P.; Tuzi, A. A Computer-Model of Molecular Arrangement in a N-Paraffinic Luquid. *J. Chem. Phys.* **1980,** 73 (1), 548-552 DOI: 10.1063/1.439853.

86. Dodd, L. R.; Boone, T. D.; Theodorou, D. N. A Concerted Rotation Algorithm for Atomistic Monte-Carlo Simulation of Polymer Melts and Glasses. *Mol. Phys.* **1993,** 78 (4), 961-996 DOI: 10.1080/00268979300100641.

87. Pant, P. V. K.; Theodorou, D. N. Variable Connectivity Method for the Atomistic Monte-Carlo Simulation of Polydisperse Polymer Melts. *Macromolecules* **1995,** 28 (21), 7224-7234 DOI: 10.1021/Ma00125a027.

**Appendix A. Acceptance criterion of the ring-flip move**

In the first step of the ring flip move, the position of atom 2 is altered by changing the rotational angle $\omega$ that defines the rotation of this atom about an azimuthal axis connecting atoms 1 and 3 (Figure A.1) with known bond lengths $l_{12}$ and $l_{23}$.

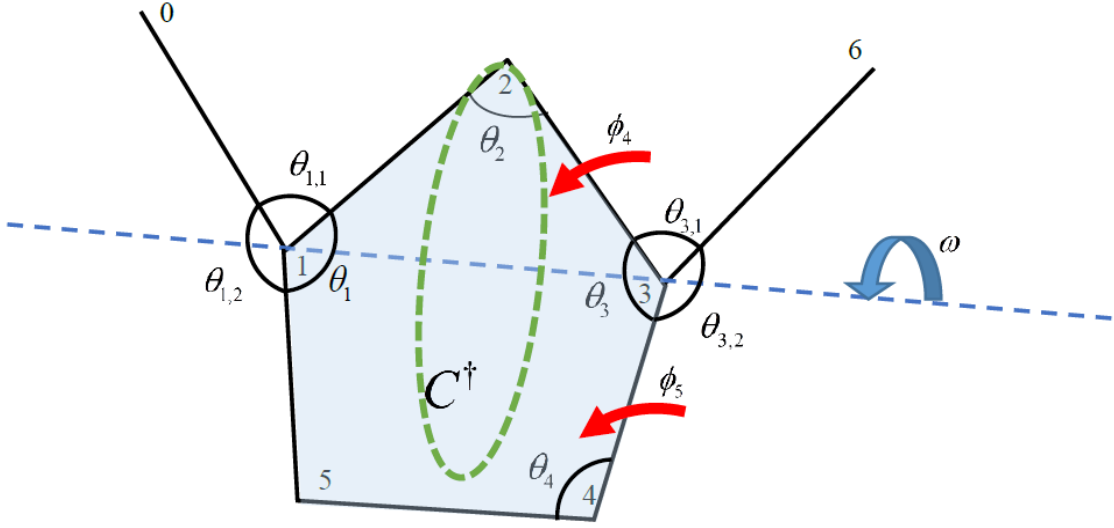

**Figure A.1.** Schematic representation of the ring-flip Monte Carlo move.

Thus, atom 2 lies on the circle $C^\dagger$ and its position can be expressed as

$$\mathbf{r}_2(x_2, y_2, z_2) \to \mathbf{r}_2(l_{12}, l_{23}, \omega) \tag{A.1}$$

In the second step of the move, the positions of atoms 4 and 5 are specified by applying the contraints of ring planarity and rigidity. In the local Flory frame of bonds 3-4 and 4-5, respectively, these can be expressed as

$$\begin{aligned}\mathbf{r}_4(x_4, y_4, z_4) &\to \mathbf{r}_4(l_{34}, \theta_3, \varphi_4) \\ \mathbf{r}_5(x_5, y_5, z_5) &\to \mathbf{r}_5(l_{45}, \theta_4, \varphi_5)\end{aligned} \tag{A.2}$$

By using the following expressions:

$$\begin{aligned}d\mathbf{r}_2 &= \frac{l_{12}l_{23}\sin^2\theta_2}{l_{12}^2 + l_{23}^2 - 2l_{12}l_{23}\cos\theta_2} dl_{12} dl_{23} d\omega \\ d\mathbf{r}_4 &= l_{34}^2 \sin\theta_3 dl_{34} d\theta_3 d\phi_4 \\ d\mathbf{r}_5 &= l_{45}^2 \sin\theta_4 dl_{45} d\theta_4 d\phi_5\end{aligned} \tag{A.3}$$

for the corresponding differential volumes (see ref 70 in the main text), the probability of the new configuration is

$$P \propto \exp(-\beta U) d\mathbf{r}_2 d\mathbf{r}_4 d\mathbf{r}_5$$
$$\propto \frac{l_{12} l_{23} \sin^2 \theta_2}{l_{12}^2 + l_{23}^2 - 2 l_{12} l_{23} \cos \theta_2} \exp(-\beta U) l_{34}^2 \sin \theta_3 l_{45}^2 \sin \theta_4 dl_{12} dl_{23} d\omega dl_{34} d\theta_3 d\phi_4 dl_{45} d\theta_4 d\phi_5 \quad (A.4)$$

where $U$ is the total potential energy of the new configuration. In our implmentation of the ring-flip move, all bond lenths, bond bending and torsion angles involved in the above expression are held fixed. Thus, we can integrate them out, leading to:

$$P \propto \exp(-\beta U) d\omega \quad (A.5)$$

Therefore, the Metropolis criterion of the ring-flip move following the implementation through a random selection of the rotational angle $\omega$ should read:

$$p_{acc} = \min\left[1, \exp(-\beta \Delta U)\right] \quad (A.6)$$

where $\Delta U$ is the total energy difference between old and new states in the attempted move.

**Appendix B. Theoretical expressions for the probability distribution of bond and torsional angles around the branch point for phantom chains**

**B.1 Probability distribution of the bond angle quad $(\theta_{2,1}, \theta_{2,2}, \theta_{3,1}, \theta_{3,2})$**

To calculate the joint probability distribution $P(\theta_{2,1}, \theta_{2,2}, \theta_{3,1}, \theta_{3,2})$ of the bond angle quad $(\theta_{2,1}, \theta_{2,2}, \theta_{3,1}, \theta_{3,2})$ we work as follows. The probability distribution $P(\theta_{2,1}, \theta_{2,2}, \theta_{3,1}, \theta_{3,2})$ for the quad of bond angles $(\theta_{2,1}, \theta_{2,2}, \theta_{3,1}, \theta_{3,2})$ involved in the construction of the 2$^{nd}$ ring (see Figure 1 in the main text) is given by the equation:

$$P(\theta_{2,1}, \theta_{2,2}, \theta_{3,1}, \theta_{3,2}) \propto \left( \prod_{i=2}^{3} \prod_{j=1}^{2} \exp\left[ -\beta U_{\text{bend}}(\theta_{i,j}) \right] \right) d^3\mathbf{r}_3 d^3\mathbf{r}_4 d^3\mathbf{r}_5 d^3\mathbf{r}_6 d^3\mathbf{r}_7 \quad \text{(B.1.1)}$$

The position vectors of the atoms can be expressed in terms of internal degrees of freedom. For atoms 4, 5 and 6 we use the length of the bond, the bond angle and the torsion angle defining their position while for atoms 3 and 7 we use the bond length and the two bond angles:

$$\mathbf{r}_3(x_3, y_3, z_3) \rightarrow \mathbf{r}_3(l_{23}, \theta_{2,1}, \theta_{2,2})$$
$$\mathbf{r}_4(x_4, y_4, z_4) \rightarrow \mathbf{r}_4(l_{34}, \theta_{3,1}, \phi_4)$$
$$\mathbf{r}_5(x_5, y_5, z_5) \rightarrow \mathbf{r}_5(l_{45}, \theta_4, \phi_5) \quad \text{(B.1.2)}$$
$$\mathbf{r}_6(x_6, y_6, z_6) \rightarrow \mathbf{r}_6(l_{56}, \theta_5, \phi_6)$$
$$\mathbf{r}_7(x_7, y_7, z_7) \rightarrow \mathbf{r}_7(l_{37}, \theta_3, \theta_{3,2})$$

Employing the Jacobians of these transformations (see ref 70 in the main text):

$$d\mathbf{r}_3 = \frac{l_{23}^2 \sin\theta_{2,1} \sin\theta_{2,2}}{\sqrt{1 - \cos^2\theta_2 - \cos^2\theta_{2,1} - \cos^2\theta_{2,2} + 2\cos\theta_2 \cos\theta_{2,1} \cos\theta_{2,2}}} dl_{23} d\theta_{2,1} d\theta_{2,2}$$

$$d\mathbf{r}_4 = l_{34}^2 \sin\theta_{3,1} dl_{34} d\theta_{3,1} d\phi_4$$

$$d\mathbf{r}_5 = l_{45}^2 \sin\theta_4 dl_{45} d\theta_4 d\phi_5 \quad \text{(B.1.3)}$$

$$d\mathbf{r}_6 = l_{56}^2 \sin\theta_5 dl_{56} d\theta_5 d\phi_6$$

$$d\mathbf{r}_7 = \frac{l_{37}^2 \sin\theta_3 \sin\theta_{3,2}}{\sqrt{1 - \cos^2\theta_3 - \cos^2\theta_{3,2} - \cos^2\theta_{3,1} + 2\cos\theta_3 \cos\theta_{3,2} \cos\theta_{3,1}}} dl_{37} d\theta_3 d\theta_{3,2}$$

and considering that all the bond lengths $(l_{23}, l_{34}, l_{45}, l_{56}, l_{37})$, the bond angles $\theta_3$, $\theta_4$ and $\theta_5$ and the two torsion angles $\phi_5$ and $\phi_6$ are fixed, equation B.1.1 becomes:

$$P(\theta_{2,1},\theta_{2,2},\theta_{3,1},\theta_{3,2}) \propto$$

$$\frac{\left(\prod_{i=2}^{3}\prod_{j=1}^{2}\exp\left[-\beta U_{bend}(\theta_{i,j})\right]\right)}{\sqrt{1-\cos^2\theta_2-\cos^2\theta_{2,1}-\cos^2\theta_{2,2}+2\cos\theta_2\cos\theta_{2,1}\cos\theta_{2,2}}}$$

$$\frac{\sin\theta_{3,1}\sin\theta_{3,2}}{\sqrt{1-\cos^2\theta_3-\cos^2\theta_{3,1}-\cos^2\theta_{3,2}+2\cos\theta_3\cos\theta_{3,1}\cos\theta_{3,2}}}d\theta_{2,1}d\theta_{2,2}d\theta_{3,1}d\theta_{3,2}d\varphi_4 \quad (B.1.4)$$

which is the desired expression for $P(\theta_{2,1},\theta_{2,2},\theta_{3,1},\theta_{3,2})$. We observe that this can be broken down to the product of two independnet probabilities, one involving the pair of atoms $(\theta_{2,1},\theta_{2,2})$ and the other the pair $(\theta_{3,1},\theta_{3,2})$. That is:

$$P(\theta_{2,1},\theta_{2,2},\theta_{3,1},\theta_{3,2}) \propto P(\theta_{2,1},\theta_{2,2})\, P(\theta_{3,1},\theta_{3,2}) \quad (B.1.5)$$

Indicating that the problem of calculating the probability of the quad $(\theta_{2,1},\theta_{2,2},\theta_{3,1},\theta_{3,2})$ is reduced to the problem of calculating the joint probability of the dyads $(\theta_{2,1},\theta_{2,2})$ and $(\theta_{3,1},\theta_{3,2})$, respectively. The two probabilities $P(\theta_{2,1},\theta_{2,2})$ and $P(\theta_{3,1},\theta_{3,2})$ have the same functional form. Therefore in Appendix B.2 we will focus only on $P(\theta_{2,1},\theta_{2,2})$.

## B.2 Probability distribution of bond angle dyad $(\theta_{2,1},\theta_{2,2})$

The probability distrubution of the bond angle dyad $(\theta_{2,1},\theta_{2,2})$ is obtained directly by integrating eq (B.1.4) with respect to $\theta_{3,1}$ and $\theta_{3,2}$:

$$P(\theta_{2,1},\theta_{2,2}) = d\theta_{2,1}d\theta_{2,2}\int d\theta_{3,1}d\theta_{3,2}\, P(\theta_{2,1},\theta_{2,2},\theta_{3,1},\theta_{3,2})$$

$$\propto \frac{\sin\theta_{2,1}\sin\theta_{2,2}\exp\left[-\beta U_{bend}(\theta_{2,1})\right]\exp\left[-\beta U_{bend}(\theta_{2,2})\right]}{\sqrt{1-\cos^2\theta_2-\cos^2\theta_{2,1}-\cos^2\theta_{2,2}+2\cos\theta_2\cos\theta_{2,1}\cos\theta_{2,2}}}d\theta_{2,1}d\theta_{2,2} \quad (B.2.1)$$

Furthermore, given that $\theta_2$ is constant, eq (B.2.1) shows that the probability distribution $P(\theta_{2,1},\theta_{2,2})$ is symmetric with respect to angles $\theta_{2,1}$ and $\theta_{2,2}$. Thus, the probability distributions $P(\theta_{2,1})$ and $P(\theta_{2,2})$ for the two angles $\theta_{2,1}$ and $\theta_{2,2}$, respectively, are given by the same functional form. This can be calculated by just integrating out in eq (B.2.1):

$$\rho(\theta_{2,1}) = \frac{\int_0^\pi d\theta_{2,2} P(\theta_{2,1}, \theta_{2,2})}{\int_0^\pi d\theta_{2,1} \int_0^\pi d\theta_{2,2} P(\theta_{2,1}, \theta_{2,2})} \tag{B.2.2a}$$

$$\rho(\theta_{2,2}) = \frac{\int_0^\pi d\theta_{2,1} P(\theta_{2,1}, \theta_{2,2})}{\int_0^\pi d\theta_{2,2} \int_0^\pi d\theta_{2,1} P(\theta_{2,1}, \theta_{2,2})} \tag{B.2.2b}$$

**B.3 Probability distribution of the interring dihedral angle $\phi_3 = \phi_{C-S-C-C}$**

The probability of dihedral angle $\phi_3$ is directly related to the probability of the position of atom 3 using the transformation $\mathbf{r}_3(x_3, y_3, z_3) \to \mathbf{r}_3(l_{23}, \theta_{2,1}, \phi_3)$:

$$\begin{aligned} P(\mathbf{r}_3) &\propto \exp\left[-\beta U_{\text{bend}}(\theta_{2,1})\right] \exp\left[-\beta U_{\text{bend}}(\theta_{2,2})\right] \exp\left[-\beta U_{\text{tor}}(\varphi_3)\right] d\mathbf{r}_3 \\ &\propto l_{23}^2 \sin(\theta_{2,1}) \exp\left[-\beta U_{\text{bend}}(\theta_{2,1})\right] \exp\left[-\beta U_{\text{bend}}(\theta_{2,2})\right] \exp\left[-\beta U_{\text{tor}}(\phi_3)\right] dl_{23} d\theta_{2,1} d\phi_3 \end{aligned} \tag{B.3.1}$$

For the phantom chain model where $U_{\text{tor}}(\phi_3) = 0$, and given that the bond length $l_{23}$ is constant, eq (B.3.1) becomes:

$$P(\mathbf{r}_3) \propto \sin\theta_{2,1} \exp\left[-\beta U_{\text{bend}}(\theta_{2,1})\right] \exp\left[-\beta U_{\text{bend}}(\theta_{2,2})\right] d\theta_{2,1} d\phi_3 \tag{B.3.2}$$

The probability of $\phi_3$ is obtained by integrating eq (B.3.4) over $\theta_{2,1}$:

$$P(\phi_3) \propto d\phi_3 \int \sin\theta_{2,1} \exp\left[-\beta U_{\text{bend}}(\theta_{2,1}) - \beta U_{\text{bend}}(\theta_{2,2})\right] d\theta_{2,1} \tag{B.3.3}$$

The bond bending angle $\theta_{2,2}$ can immediately be calculated from the dot product of the vectors $\mathbf{r}_{23}$ and $\mathbf{r}_{28}$ as function of $\theta_{2,1}$ and $\phi_3$. Expressing the positions of atoms 3 and 8 in the local Flory frame of bonds 2-3 and 2-8 (see Figure 1 in the main text) we have that

$$\mathbf{r}_{23} = \left[-l_{23}\cos\theta_{2,1}, l_{23}\sin\theta_{2,1}\cos\phi_3, l_{23}\sin\theta_{2,1}\sin\phi_3\right] \tag{B.3.4}$$

$$\mathbf{r}_{28} = \left[-l_{28}\cos\theta_2, l_{28}\sin\theta_2, 0\right] \tag{B.3.5}$$

and finally:

$$\theta_{2,2} = \cos^{-1}\left(\frac{\mathbf{r}_{23}\cdot\mathbf{r}_{28}}{l_{23}l_{28}}\right) = \cos^{-1}\left[\cos\theta_2\cos\theta_{2,1} + \sin\theta_2\sin\theta_{2,1}\cos\phi_3\right] \tag{B.3.6}$$

The corresponding normalized probability density of $\phi_3$ is then

$$\rho(\phi_3) = \frac{\int_0^\pi \sin\theta_{2,1} \exp\left[-\beta U_{bend}(\theta_{2,1}) - \beta U_{bend}(\theta_{2,2})\right] d\theta_{2,1}}{\int_0^{2\pi} d\phi_3 \int_0^\pi \sin\theta_{2,1} \exp\left[-\beta U_{bend}(\theta_{2,1}) - \beta U_{bend}(\theta_{2,2})\right] d\theta_{2,1}} \tag{B.3.7}$$

**Monte Carlo algorithm based on internal bridging moves for the atomistic simulation of thiophene oligomers and polymers**

Flora D. Tsourtou, Stavros D. Peroukidis, Loukas D. Peristeras, and Vlasis G. Mavrantzas

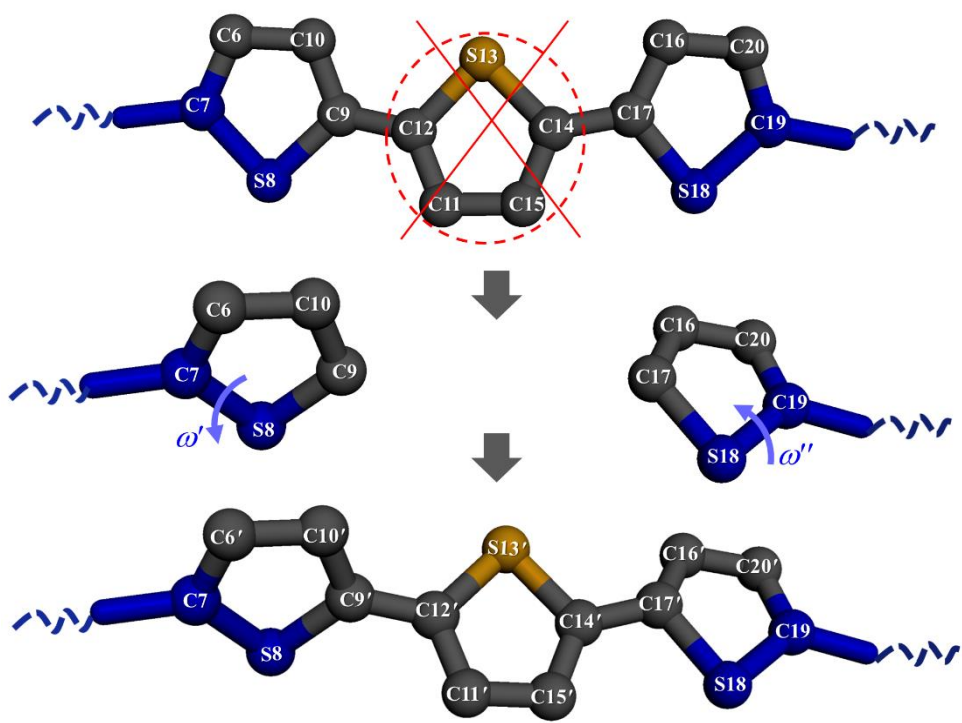